\documentclass{IEEEtran}
\usepackage{array}
\usepackage{graphicx, subfigure}
\usepackage{float}
\usepackage{multirow}
\usepackage{rotating}
\usepackage{algorithmic,algorithm}
\usepackage{amsmath} 
\usepackage{amsfonts}
\usepackage{amssymb}  
\usepackage{amsthm}
\usepackage{bm}
\usepackage{color}

\newtheorem{lemma}{Lemma}
\def\N{{\mathcal{N}}}

\begin{document}
\title{A control theoretic approach to achieve proportional fairness in 802.11e EDCA WLANs }
\author{Xiaomin Chen, Ibukunoluwa Akinyemi and Shuang-Hua Yang\IEEEauthorrefmark{1}
\\Department of Computer Science, Loughborough University
\thanks{\IEEEauthorrefmark{1} corresponding author S.H.Yang@lboro.ac.uk}}
\maketitle

\begin{abstract}
This paper considers proportional fairness amongst ACs in an EDCA WLAN {for} provision of distinct QoS requirements and priority parameters. A detailed theoretical analysis is provided to derive the optimal station attempt probability which leads to a proportional fair allocation of station throughputs. The desirable fairness can be achieved using a centralised adaptive control approach. This approach is based on multivariable state-space control theory and uses the Linear Quadratic Integral (LQI) controller to periodically update $CW_{min}$ till the optimal fair point of operation.  Performance evaluation demonstrates that the control approach has high accuracy performance and fast convergence speed for general network scenarios.  To our knowledge this might be the first time that a closed-loop control system is designed for EDCA WLANs to achieve proportional fairness.
\end{abstract}

\section{Introduction}

Enhanced Distributed Channel Access (EDCA) was proposed in IEEE 802.11e-2005 standard to support QoS enhancement and service differentiation for WLAN applications~\cite{80211estd}. It extends the basic Distributed Coordination Function (DCF) by classifying traffic flows into four different Access Categories (AC), namely voice, video, best-effort and background.  Traffic with higher QoS requirements, e.g. shorter delay deadline,  is assigned {a} higher priority, and hence, on average, waits for less time before being sent to the channel. This mechanism is beneficial for high-priority traffic. Compared to the DCF, EDCA sacrifices the performance of low-priority traffic to some extent to provide QoS support for high-priority traffic. When the network is saturated with a large proportion of high-priority flows, an extremely unfair scenario will appear, in which the channel will be almost completely occupied by high-priority flows, e.g. VoIP or video streaming flows, however low-priority traffic, such as email or web browsing data, will suffer severe starvation.

Resource allocation in EDCA WLANs has therefore been the subject of considerable interest. The objective is to seek for a fair allocation of network resources (e.g. throughput, airtime and etc.) amongst different traffic types, and meanwhile, guarantee the specific QoS requirements and service differentiation. This paper considers proportional fair allocation of station throughputs amongst ACs {for} provision of distinct average delay deadlines and priority parameters. The 802.11e EDCA standard specifies four contention parameters to distinguish priority levels, which are minimum Contention Window ($CW_{min}$), maximum Contention Window ($CW_{max}$), Arbitration Inter Frame Space ($AIFS$) and  maximum Transmission Opportunity ($TXOP$). A set of default values for the four parameters  are recommended in the standard for each physical (PHY) layer supported by 802.11e. As the default values do not take into account the varying WLAN conditions, and thus lead to suboptimal performance and no fairness guarantees, in this paper we find the optimal $CW_{min}$ value that leads to proportional fair allocation of station throughputs while assuming  $AIFS$ and $TXOP$ taking the recommended values and $CW_{max}=CW_{min}$. The optimal $CW_{min}$ value corresponds to an optimal station attempt probability which is derived from the proportional fairness analysis.

In order to implement the derived proportional fair allocation in practice, a centralised adaptive approach which uses multivariable state-space control theory is then proposed.
The WLAN is represented as a discrete multi-input multi-output (MIMO) linear time-invariant (LTI) state-space model. A state feedback control method, the Linear Quadratic Integral (LQI) control, is used to tune the $CW_{min}$ value to drive the station attempt probability to the optimum so as to maintain a fair throughput allocation. We have demonstrated in simulations that the proposed control approach is adaptive to general network scenarios with high accuracy and fast convergence speed. To our knowledge this might be the first time that a closed-loop control system is designed for EDCA WLANs to achieve proportional fairness amongst ACs.

The remainder of this paper is organised as follows. {Section~\ref{Sec:RelatedWork} gives a comprehensive review {\color{black} of} the state-of-the-art research on fairness and control theory approaches to solve network problems.} Section~\ref{Sec:Fairness} presents the theoretical analysis to derive the optimal station attempt probability which leads to proportional fair allocation of station throughputs given the constraints on the average delay deadlines for different ACs. Section~\ref{Sec:ControlAlg} describes a centralised adaptive control approach which can realise the proportional fair allocation derived from Section~\ref{Sec:Fairness} in real networks. Section~\ref{Sec:Simulation} evaluates the performances of the fairness algorithm and the proposed centralised control approach, and Section~\ref{Sec:Conclusion} concludes the paper.

\section{State of the art}\label{Sec:RelatedWork}
\subsection{Fairness}
Fairness has been the subject of a considerable body of literature on 802.11 WLANs~\cite{Bharghavan}-\cite{MaxminDoug}. The unfairness behaviors may be caused by a number of factors, e.g. hidden terminals, exposed terminals, capture, uplink/downlink unfairness, asymmetric radio conditions and multiple data rates and etc., which have been investigated in~\cite{Bharghavan, Jiang, Kochut, TCP, Wang, Malone, Albert, Siris}. There also exist distinct fairness criteria that are widely adopted in network resource allocation, such as time-based fairness, throughput-based fairness, proportional fairness, max-min fairness, weighted fairness and etc. (see for example \cite{Albert, Alex, Xiaomin, MaxminDoug, Kelly, Li, Siris, Pablo} and references therein). In this paper, we employ the proportional fairness criterion to deal with the unfairness amongst ACs with differentiated priority parameters.

The CSMA/CA scheduling used in 802.11 differs fundamentally from wired networks due to carrier sense
deferral of the contention window countdown and the occurrence
of colliding transmissions, both of which act to couple
together the scheduling of station transmissions and lead to the rate region being nonconvex~\cite{nonconvex}. Therefore, well established utility fairness
techniques from wired and TDMA networks cannot be directly
applied to random access CSMA/CA wireless networks, and hence most studies on proportional fairness in 802.11 WLANs are confined to approximation approaches~\cite{Bharghavan,Dunn,Siris}. \cite{Alex} corrects prior
approximate studies and provides the first rigorous analysis of proportional
fairness in 802.11 WLANs. It shows that there exists a unique proportional
fair rate allocation and completely characterises the
allocation in terms of the total air-time quantity. \cite{Xiaomin} extends the work in \cite{Alex} by considering lossy links and BSC-based coding with delay deadline constraints. The optimal joint allocation
of airtime and coding rate allows the throughput/loss/delay trade–off amongst flows sharing network resources to be performed in a principled manner. The proportional fairness analysis in this paper builds upon the approaches used in \cite{Xiaomin} and \cite{Prem}.

 In particular, the fairness issue in 802.11e EDCA WLANs has been given considerable attention.  \cite{Albert} derives a throughput allocation based on the proportional fair criterion in multirate 802.11e WLANs. It shows that in a proportional fair allocation high and low bit rate stations are assigned with the same share of channel time, and thus high bit rate stations obtain higher throughput. Two schemes that respectively involve Contention Window $CW$ and Transmission Length $TL$ (which is based on $TXOP$) are then proposed to achieve this allocation.  \cite{Siris} investigates the weighted proportional fairness in both single-rate and multi-rate 802.11e WLANs via test-bed experiments, and compares proportional fairness with time-based fairness in a multi-rate setting. Proportional fairness can be achieved by tailoring the $CW_{min}$, and the time-based fairness can be achieved by adjusting packet size or $TXOP$  limit. It concludes that in a multi-rate 802.11e WLAN proportional fairness with equal weights achieves higher performance than time-based fairness in terms of both aggregate utility and throughput. \cite{Albert} and \cite{Siris} deal with the unfairness behaviour arising in a WLAN due to asymmetric channel conditions, and \cite{Siris} investigates the problem through test-bed evaluations relying on no theoretical analysis foundation. \cite{Lee, Pablo, Cheng} address the unfairness problem existing amongst ACs. The priority-based service supported by 802.11e EDCA, while allowing differentiated service for flows of different priorities, cannot ensure service amount in proportion to their demands. \cite{Lee} proposes a mechanism called Weighted Fair-EDCA (WF-EDCA) which uses Distributed Fair Scheduling (DFS) in each backoff entity to provide weighted proportional fair service among different ACs.  \cite{Pablo} proposes an algorithm to compute the optimal configuration of the EDCA differentiation parameters given a set of QoS requirements in terms of throughput and delay  with multiple real-time and data ACs. A throughput and delay analysis  is provided, based upon which the optimal configuration algorithm is derived by maximising the throughput using the weighted max-min fairness criterion. {\cite{Cheng} presents a new scheme which exploits differentiations of both inter frame space and contention window  to achieve weighted fairness for two classes of services under EDCA mode in an 802.11e WLAN. Given the AIFSs, the proposed scheme can properly set the corresponding CWs such that the ratio of the two classes' successful transmission probabilities can attain a pre-defined weighted-fairness goal. \cite{Abeysekera} proposes a dynamic contention window control scheme to achieve fairness between uplink and downlink TCP flows for the IEEE 802.11e EDCA-based WLANs while guaranteeing QoS requirements for real-time traffic. The proposed scheme first determines the minimum contention window size in the best-effort access category at APs, and then determines the minimum and maximum contention window sizes in higher priority access categories, such as voice and video, so as to guarantee QoS requirements for these real-time traffic.} In this paper, we provide a comprehensive throughput and delay analysis that incorporates EDCA differentiation parameters for 802.11e EDCA WLANs and derive a throughput allocation to achieve proportional fairness amongst ACs.

\subsection{Control theory approach}

Control theory has been applied to the area of communication networks in a wide range of aspects. For instance, \cite analyses a combined TCP and Active Queue Management (AQM) model
from a control theoretic standpoint. It uses a nonlinear dynamic
model of TCP to design a feedback control system depiction of AQM using the random early detection (RED) scheme. \cite{Mario} introduces a control theoretical analysis
of the closed-loop congestion control problem in packet networks. The control theoretical approach is used in a proportional rate controller, where packets are admitted into the network in
accordance with network buffer occupancy. A Smith Predictor is
used to deal with large propagation delays. \cite{Park} proposes a QoS-provisioning feedback control framework in order to achieve TCP uplink/downlink fairness and service differentiation. The medium access price (MAP) is delivered to TCP senders and the TCP senders adjust their sending rates to reduce congestion at the interface queue of the home gateway in an 802.11-based home network. \cite{Paul1} proposes a centralised adaptive control (CAC) approach to dynamically adjust the $CW_{min}$ configuration of 802.11 WLANs with the goal of minimising the overall throughput performance. A proportional integrator (PI) controller is used to establish a closed-loop control system. \cite{Paul2} extends the work in \cite{Paul1} by considering real-time traffic in 802.11e WLANs. The $CW_{min}$ configuration is adjusted in terms of minimising the average delay, which results in a better quality of experience (QoE) of the video traffic. \cite{Paul3} proposes a distributed adaptive control (DAC) algorithm based on the multivariable control theory. An independent PI controller is installed at each station in a WLAN and uses locally available information to drive the overall network performance to the optimum. In addition to the closed-loop control methods,  \cite{Bio} proposes an open-loop self-adaptive  rate control  approach for multi-priority WLANs. The approach is based on a biological competitive model which guides data flows to compete for network bandwidth in the way of a native ecosystem. The model parameters self-tune themselves to optimise the bandwidth utilization in an EDCA WLAN with multiple ACs.   In this paper we employ the multivariable feedback control method to design a centralised adaptive control approach to achieve proportional fairness amongst different ACs. The WLAN is represented as a discrete MIMO LTI state-space model. The state feedback LQI controller is used to tune the $CW_{min}$ value. To the best of our knowledge, this might be the first time that a closed-loop control system is used in 802.11 WLANs to deal with fairness issue.

\section{Proportional fairness in multi-priority 802.11e WLANs}\label{Sec:Fairness}

\subsection{Network model}

\begin{figure}
  \begin{center}
  \includegraphics[width=0.48\textwidth]{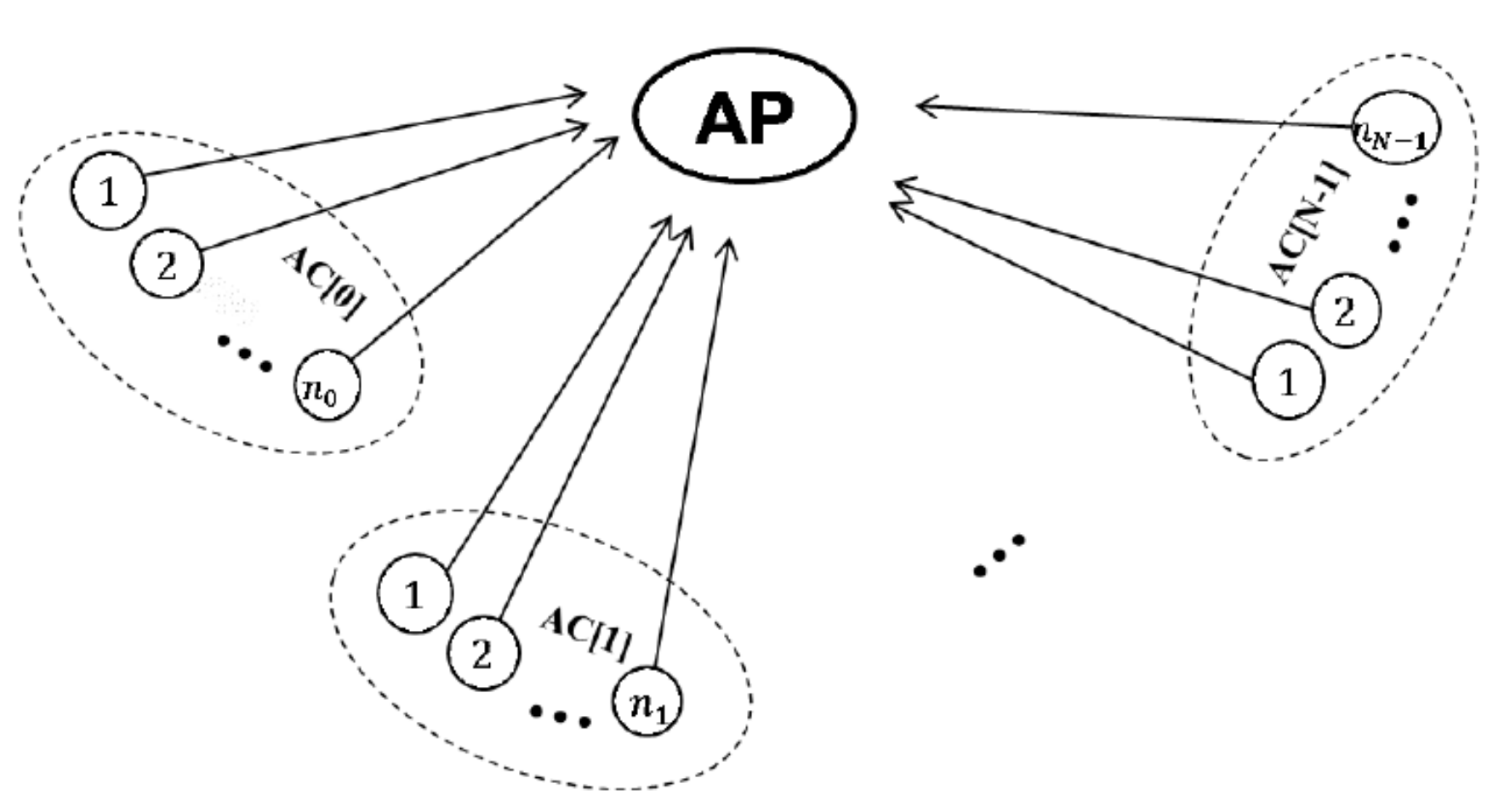}\\
   \caption{ Network model. }\label{Fig:Model}
  \end{center}
\end{figure}

We consider a single-hop 802.11e EDCA WLAN with one AP and $n$ client stations, as depicted in Fig.~\ref{Fig:Model}. The channel is assumed to be error-free for all supported PHY rates. Traffic flows are classified into $N$ different ACs. We assume that each client carries flows of a single AC, so there are no virtual collisions in our setup. The number of stations in the $i$th AC is $n_i$.
The total number of stations is thus $n=\sum_{i=0}^{N-1}n_i$.   The analysis can be readily generalised to
encompass situations where client stations have lossy links and carry more than one AC. But the current simplified error-free model is sufficient to capture performance features of differentiation settings in WLANs.
\subsection{Station throughput}\label{SubSec:throughput}
We start with the analysis of station throughput under saturation conditions, {\color{black}i.e. there is always a packet waiting to be transmitted in the queue of a station}.
For the $i$th AC, the following parameters are defined: $CW_{min}^i$ is the minimum contention window; $CW_{max}^i$ is the maximum contention window; $t_i$ is the number of time slots in the period of $AIFS_i$, i.e. $AIFS_i=SIFS+t_i\times\sigma$ where $SIFS$ is the duration of the Short Interframe Space and $\sigma$ represents the duration of a physical time slot;  $m_i$ is the number of packets transmitted in a TXOP burst. We assume that packets of all ACs have the same length of $L$ bits, and are transmitted at the same PHY rate $r$ Mbps under the assumption of error-free channels.  Due to the use of TXOP bursting, the RTS/CTS exchange mechanism is used to make fast recovery from collisions. The notation used in this paper {\color{black}is} listed in Table.~\ref{Notation}.
\begin{table}
\caption{Notation}\label{Notation}
\centering
\begin{tabular}{|c|c|}
  \hline
    $N$                & Number of ACs\\
    $n_i$              & Number of STAs in AC $i$ \\
    $n$                & {Total} number of STAs \\
    $CW_{min}^i$, $W_i$      & Minimum contention window of AC $i$ \\
    $CW_{max}^i$       & Maximum contention window of AC $i$ \\
    $AIFS_i$           & Duration of AIFS of  AC $i$ \\
    $SIFS$             & Duration of SIFS \\
    $EIFS$             & Duration of EIFS \\
    $T_{RTS}$          & Duration of RTS \\
     $T_{CTS}$          & Duration of CTS \\
    $\sigma$           & Duration of a physical time slot\\
    $t_i$              & Number of time slots in $AIFS_i$ \\
    $t_{min}$          & Minimum $t$ value among all ACs \\
    $m_i$             & Number of packets in the TXOP burst of AC $i$ \\
    $L$              &  Packet size \\
    $r$               &  PHY data rate \\
    $\tau_i$          & Station attempt probability of AC $i$ \\
    $P^{idle}$        & Slot idle probability \\
    $P_i^{succ}$      & Probability of a successful transmission by a station of AC $i$ \\
    $P^{succ}$        & Probability of a successful transmission in a time slot \\
    $s_i$             & Station throughput of AC $i$ \\
    $T^{col}$   & Duration of a collision \\
    $T_i^{succ}$ & Duration of a successful transmission from a station of AC $i$ \\
    $T_i^o$   & Protocol {\color{black}overhead} of a TXOP burst  of AC $i$ \\
    $T^{oo}$  & Protocol {\color{black}overhead} of a single packet within TXOP burst \\
    $M$ & Retry limit in exponential backoff algorithm \\
    $P_i^{col}$ & Conditional collision probability of AC $i$ \\
    $P_i^{blk}$ & Blocking probability of AC $i$ \\
    $D_i^{cd}$ & Expected countdown delay of AC $i$ \\
     $D_i^{blk}$ & Expected blocking delay of AC $i$ \\
     $D_i^{retx}$ & Expected retransmission delay of AC $i$ \\
     $D_i^{succ}$ & Expected retransmission delay of AC $i$  \\
     $D_i$   & Average delay of a TXOP burst of AC $i$ \\
     $d_i$ & Delay deadline of a single packet of AC $i$ \\
     $\boldsymbol{H}$ & Plant lumbarisation matrix \\
     $\boldsymbol{x(k)}$ & System state at instant $k$ \\
     $\boldsymbol{y(k)}$ & System output at instant $k$ \\
     $\boldsymbol{u(k)}$ & System input  at instant $k$ \\
       $\boldsymbol{r(k)}$ & Controller input  at instant $k$ \\
       $\boldsymbol{K}$ & LQI optimal gain matrix  \\
        $\boldsymbol{Q}$ & LQI state cost weighting matrix  \\
         $\boldsymbol{R}$ & LQI control cost weighting matrix  \\

     \hline
 \end{tabular}
\end{table}

A MAC time slot may either be a PHY idle slot, a successful transmission or a colliding transmission. Let $\tau_i$ denote the probability that a station carrying a flow of AC $i$ attempts to transmit in a time slot, so $0 <\tau_i<1$. The probability that a time slot is idle is
\begin{equation*}
P^{idle}=\prod_{i=0}^{N-1}(1-\tau_i)^{n_i}
\end{equation*}
As the channel is assumed to be error-free, packet losses are only caused by collisions. The probability that a station with a flow of AC $i$ makes a successful transmission is then
\begin{equation*}\begin{split}
P_i^{succ}=\tau_{i}(1-\tau_i)^{n_i-1}\prod_{j=0,j\neq i}^{N-1}(1-\tau_j)^{n_j}
=\frac{\tau_{i}}{1-\tau_{i}}P^{idle}
\end{split}
\end{equation*}
The probability that a time slot is a successful transmission is then
\begin{equation*}
P^{succ}=\sum_{i=0}^{N-1}n_iP_i^{succ}=P^{idle}\sum_{i=0}^{N-1}\frac{n_i\tau_{i}}{1-\tau_{i}}
\end{equation*}
The throughput of a station carrying a flow of AC $i$ is thus given by
\begin{equation*}
s_{i}(\boldsymbol{\tau})=\frac{P_i^{succ}m_iL}{{P^{idle}}\sigma+\sum\limits_{i=0}^{N-1}n_iP_i^{succ}T_i^{succ}+(1-P^{idle}-P^{succ})T^{col}}
\end{equation*}
in which $T^{col} = T_{RTS}+EIFS$ is the duration of a collision. $EIFS$ represents the duration of the Extended Interframe Space used in 802.11 WLANs, {which is given by $EIFS=T_{ACK} + SIFS + DIFS$}.  Note that $T^{col}$ is the same for all ACs due to the use of RTS/CTS handshaking. $T_i^{succ}$ is the duration of a successful transmission from a station sending traffic of AC $i$. It depends on the size of TXOP packet burst, and is thus given by
\begin{equation*}
  T_i^{succ}=T_i^{o}+m_i(T^{oo}+\frac{L}{r})
\end{equation*}
where $T_i^o=T_{RTS}+SIFS+T_{CTS}+AIFS_i$ is the protocol {\color{black}overhead} associated with the transmission of a TXOP burst; $T^{oo}$ is the protocol {\color{black}overhead} associated with each packet transmission within a TXOP burst, i.e. $T^{oo}=T_{PHYhdr}+2SIFS+T_{ACK}$.

By working in terms of the quantity $\alpha_i=\frac{\tau_i}{1-\tau_i}, \alpha_i>0$ instead of $\tau_i$, the station throughput is rewritten as
\begin{equation}\label{thrpt}
s_{i}=\frac{\alpha_{i}m_iL}{X\cdot T^{col}}
\end{equation}
in which
\begin{equation*}
X=\frac{\sigma}{T^{col}}+\sum\limits_{i=0}^{N-1}n_i\Big(\frac{T_i^{succ}}{T^{col}}-1\Big)\alpha_{i}+\prod\limits_{i=0}^{N-1}\big(1+\alpha_{i}\big)^{n_i}-1
\end{equation*}

\subsection{Average delay}\label{SubSec:delay}
Next, we will calculate the average delay experienced by a TXOP burst of each AC. {We start with the analysis for ordinary 802.11 MAC scheduling with binary exponential backoff algorithm and then move onto the scenario when ${CW}_{min}={CW}_{max}$.}

The delay is defined in this work as the duration since a station starts contending for the medium until the transmission is finished (either received successfully or dropped because of reaching the maximum retry limit).
The calculation is based on the EDCA WLAN throughput model derived in~\cite{Kosek}.
The average delay consists of four expected delays, described as follows:
\begin{itemize}
  \item \emph {Expected countdown delay}:
   For each backoff stage $g$ ($0\leq g\leq M$, and $M$ is the retry limit. We assume that $CW_{max}^i\geq 2^M{CW}_{min}^i$), the average countdown delay for AC $i$ is $CW_{i,g}\sigma/2$, in which $CW_{i,g}=2^g{CW}_{min}^i$ is the contention window at the $g$th backoff stage. The expected delay associated with the backoff countdown process is then given by
   \begin{equation*}\begin{split}
     &D_i^{cd}=\sigma\times\bigg(\sum\limits_{g=0}^{M}(P_i^{col})^g(1-P_i^{col})\sum\limits_{h=0}^{g}\frac{CW_{i,h}}{2}\\&+(P_i^{col})^{M+1}\sum\limits_{h=0}^{M}\frac{CW_{i,h}}{2}\bigg)
   \end{split}\end{equation*}
   where $P_i^{col}$ is the conditional collision probability for the $i$th AC in the throughput model, i.e.
   \begin{equation}\label{popt}
     P_i^{col}=1-(1-\tau_i)^{n_i-1}\prod\limits_{\substack{j=0\\j\neq i}}^{N-1}(1-\tau_j)^{n_j}
   \end{equation}
   \item \emph{Expected blocking delay}: During the countdown process, when a transmission is detected on the channel the backoff time counter is ``frozen'', and reactivated again after the channel is sensed idle for a certain period. A station is called blocked in our analysis when it senses (an) ongoing transmission(s) from some other station(s) during its countdown process. The blocking delay is the period during which a station is ``frozen''. The expected number of time slots in the backoff countdown process is ${D_i^{cd}}/{\sigma}$. At each time slot, a station could be blocked by either a successful transmission or a collision. For a station of AC $i$, the delay caused by a successful transmission from some other station is
       \begin{equation*}\begin{split}
         &D_i^{bs}=T_i^{succ}(n_i-1)\tau_i(1-\tau_i)^{n_i-2}\prod_{\substack{j=0\\j\neq i}}^{N-1}(1-\tau_j)^{n_j}+\\
         &\sum_{\substack{j=0\\j\neq i}}^{N-1}T_j^{succ}n_j\tau_j(1-\tau_j)^{n_j-1}\prod_{\substack{k=0\\k\neq j,i}}^{N-1}(1-\tau_k)^{n_k}(1-\tau_i)^{n_i-1}
       \end{split}\end{equation*}
       The blocking delay because of a collision is
       \begin{equation*}\begin{split}
         &D_i^{bc}=T^{col}\bigg(1-(1-\tau_i)^{n_i-1}\prod\limits_{\substack{j=0\\j\neq i}}^{N-1}(1-\tau_j)^{n_j}\\&-(n_i-1)\tau_i(1-\tau_i)^{n_i-2}\prod_{\substack{j=0\\j\neq i}}^{N-1}(1-\tau_j)^{n_j}\\&-\sum_{\substack{j=0\\j\neq i}}^{N-1}n_j\tau_j(1-\tau_j)^{n_j-1}\prod_{\substack{k=0\\k\neq j,i}}^{N-1}(1-\tau_k)^{n_k}(1-\tau_i)^{n_i-1}\bigg)
       \end{split}\end{equation*}
       The expected blocking delay of AC $i$ is thus
       \begin{equation*}
         D_i^{blk}=\frac{D_i^{cd}}{\sigma}(D_i^{bs}+D_i^{bc})
       \end{equation*}
   \item \emph{Expected retransmission delay}: The expected retransmission delay for AC $i$ is calculated by multiplying the expected number of retransmission attempts by the collision duration, i.e.
       \begin{equation*}\begin{split}
         &D_i^{retx}=T^{col}\times\\
         &\bigg(\sum\limits_{g=0}^{M}g(P_i^{col})^g(1-P_i^{col})+(M+1)(P_i^{col})^{M+1}\bigg)
       \end{split}\end{equation*}
   \item \emph{Expected successful transmission delay}:
   The expected successful transmission delay is the duration of a successful transmission multiplied by the probability that the transmission is not dropped, which is given by
   \begin{equation*}
     D_i^{succ}=T_i^{succ}(1-(P_i^{col})^{M+1})
   \end{equation*}
\end{itemize}
Combining the above four delays, the average delay of a TXOP burst of AC $i$ is therefore given by
\begin{equation*}
  D_i=D_i^{cd}+D_i^{blk}+D_i^{retx}+D_i^{succ}
\end{equation*}

The proposed approach in this paper works by finding the optimal contention window to achieve proportional fairness amongst ACs , so the exponential backoff algorithm is unnecessary in our setting and we simply set $CW_{max}^i=CW_{min}^i$, i.e. $M=0$. To simplify notations, we hereafter refer to $CW_{min}^i$ with $W_i$. The four expected delays then become:
\begin{equation*}
  D_i^{cd}=\sigma\frac{W_i}{2}
\end{equation*}
\begin{equation*}
  D_i^{blk}=\frac{W_i}{2}(D_i^{bs}+D_i^{bc})
\end{equation*}
\begin{equation*}
  D_i^{retx}=T^{col}P_i^{col}
\end{equation*}
\begin{equation*}
  D_i^{succ}=T_i^{succ}(1-P_i^{col})
\end{equation*}

 According to the throughput model in~\cite{Kosek}, when $CW_{max}=CW_{min}$ the station attempt probability under saturation conditions can be reduced to
\begin{equation}\label{CW}
  \tau_i=\frac{2(1-P_i^{blk})}{2(1-P_i^{blk})+W_i-1}
\end{equation}
in which
\begin{equation*}
  P_i^{blk}=1-\Big[(1-\tau_i)^{n_i-1}\prod\limits_{\substack{j=0\\j\neq i}}^{N-1}(1-\tau_j)^{n_j}\Big]^{t_i-t_{min}+1}
\end{equation*}
is the probability that the backoff counter is suspended due to a busy channel during the period of $AIFS_i$. $t_{min}$ is the minimum $t$ value among all ACs.

Similarly by working in terms of the quantity $\alpha_i=\frac{\tau_i}{1-\tau_i}$,
 the average delay experienced by a TXOP burst of AC $i$ when $M=0$ is then
\begin{equation}\label{delay}\begin{split}
 & D_i=\frac{W_i(\sigma+T^{col})}{2}+\frac{Y_i}{(1+\alpha_i)^{n_i-1}}(T_i^{succ}-T^{col})+T^{col}\\&+\frac{W_iY_i}{2(1+\alpha_i)^{n_i-1}}\Big(Z_i-T^{col}+(T_i^{succ}-T^{col})(n_i-1)\alpha_i\Big)
\end{split}\end{equation}
in which
\begin{equation*}
  Y_i=\sum\limits_{j=0,j\neq i}^{N-1}{(1+\alpha_j)^{-n_j}}
\end{equation*}
\begin{equation*}
  Z_i=\sum\limits_{j=0,j\neq i}^{N-1}(T_j^{succ}-T^{col})n_j\alpha_j
\end{equation*}
and
\begin{equation*}
  W_i=\frac{2}{\alpha_i}\Big((1+\alpha_i)\prod\limits_{j=0}^{N-1}(1+\alpha_j)^{-n_j}\Big)^{t_i-t_{min}+1}+1
\end{equation*}

\subsection{Proportional fair allocation}\label{SubSec:fairness}

The 802.11e EDCA standard provides service differentiation by assigning different contention parameters to distinct ACs. Delay-sensitive traffic flows, such as voice over WLANs and streaming multimedia, are assigned with higher priorities. This mechanism has a significant cost for lower priority traffic flows as they can practically starve in dense network deployment. In this section we aim at finding the optimal ${\boldsymbol{\alpha}} := [\alpha_i]_{i\in\{0,1,\cdots,N-1\}}$ to achieve fair allocation of station throughputs amongst ACs. Meanwhile we take into account the delay constraints for each AC. The utility function is defined as the sum of the log of station throughputs，
\begin{equation*}\begin{split}
&\underset{\boldsymbol{\alpha}}
{\text{max}}\quad U(\boldsymbol{\alpha}):= \sum\limits_{i=0}^{N-1}n_i\log s_i(\boldsymbol{\alpha})\\
&\text{s. t. } \quad D_i(\boldsymbol{\alpha})\leq m_id_i \quad 0\leq i \leq N-1,\\
&\quad \quad \quad \alpha_i > 0 \quad \quad \quad \quad \ \ 0\leq i \leq N-1.
\end{split}
\end{equation*}
in which the station throughput is given by Eqn.~(\ref{thrpt}); the average delay is given by Eqn.~(\ref{delay}); $d_i$ is the delay deadline
for a single packet in a TXOP burst of AC $i$.

By plugging in the station throughput expression and removing the constant terms, the optimisation problem is simplified as
\begin{equation*}\begin{split}
&\underset{\boldsymbol{\alpha}}
{\text{max}}\quad U'(\boldsymbol{\alpha}):= \sum\limits_{i=0}^{N-1}n_i(\log \alpha_i-\log X)\\
&\text{s. t. } \quad D_i(\boldsymbol{\alpha})\leq m_id_i \quad 0\leq i \leq N-1,\\
&\quad \quad \quad \alpha_i > 0 \quad \quad \quad \quad \ \ 0\leq i \leq N-1.
\end{split}
\end{equation*}

\subsubsection{Non-convexity}
It can be verified by inspection of the second derivative that the objective function is not concave in $\boldsymbol{\alpha}$ and hence the maximisation problem is not a standard convex optimisation task. We proceed by making the log transformation $\eta_i=\log \alpha_i$. The optimisation problem then becomes
\begin{equation}\label{opt}\begin{split}
&\underset{{\boldsymbol{\eta}}}
{\text{max}}\quad U_1({\boldsymbol{\eta}}):= U'(e^{{\boldsymbol{\eta}}})\\
&\text{s. t. } \quad D_i(e^{{\boldsymbol{\eta}}})\leq m_id_i \qquad 0\leq i \leq N-1.
\end{split}
\end{equation}

\begin{lemma}\label{lem:joint_convex}
$U_1(\boldsymbol{\eta})$ is concave in $\boldsymbol{\eta}$.
\end{lemma}
The proof of Lemma 1 is included in the Appendix.

\subsubsection{Solving the optimisation with KKT conditions}
To solve this problem, we will use  Karush–Kuhn–Tucker (KKT) conditions. The Lagrangian is
\begin{equation*}
L=U_1(\boldsymbol{\eta})-\sum_{i=0}^{N-1}\mu_i\left(D_i(e^{{\boldsymbol{\eta}}})-m_id_i\right)
\end{equation*}
where the Lagrange multiplier $\mu_i\geq 0$. Differentiating the Lagrangian with respect to $\eta_i$  and setting it equal to zero yields
\begin{equation}\label{Lagrange}\begin{split}
f_i&(\boldsymbol{\alpha},\boldsymbol{\mu})=\frac{n}{X}\frac{\partial X}{\partial \eta_i}+\mu_i\frac{\partial D_i}{\partial \eta_i}+\sum\limits_{\substack{j=0\\j\neq i}}^{N-1}\mu_j\frac{\partial D_j}{\partial \eta_i}-n_i=0\\
&i\in\{0,1,\cdots,N-1\}
\end{split}\end{equation}
in which
\begin{equation*}
  \frac{\partial X}{\partial \eta_i}=\bigg(\frac{T_i^{succ}}{T^{col}}-1\bigg)n_i\alpha_i+n_i\prod_{j=0}^{N-1}(1+\alpha_j)^{n_j}\frac{\alpha_i}{1+\alpha_i}
\end{equation*}
\begin{equation*}\begin{split}
&\frac{\partial D_i}{\partial \eta_i}=\Big(\frac{Y_i\alpha_i\big((T_i^{succ}-T^{col})(n_i-1)\alpha_i+Z_i-T^{col}\big)}{2(1+\alpha_i)^{n_i-1}}\\&+\frac{(\sigma+T^{col})\alpha_i}{2}\Big)\cdot\frac{\partial W_i}{\partial \alpha_i}+\frac{Y_iW_i\alpha_i}{2}\left(1+\alpha_i\right)^{-n_i}\left(n_i-1\right)\\&\Big((T_i^{succ}-T^{col})\big((2-n_i)\alpha_i+1\big)-Z_i+T^{col}\Big)\\&+Y_i\alpha_i(T_i^{succ}-T^{col})(1-n_i)(1+\alpha_i)^{-n_i}
\end{split}\end{equation*}
\begin{equation*}\begin{split}
&\frac{\partial D_j}{\partial \eta_i}=\Big(\frac{Y_j\alpha_i\big((T_j^{succ}-T^{col})(n_j-1)\alpha_j+Z_j-T^{col}\big)}{2(1+\alpha_j)^{n_j-1}}\\&+\frac{(\sigma+T^{col})\alpha_i}{2}\Big)\cdot\frac{\partial W_j}{\partial \alpha_i}+\alpha_i{(1+\alpha_j)^{1-n_j}}\Big((T_j^{succ}-T^{col})\\&\big(\frac{W_j}{2}(n_j-1)\alpha_j+1\big)+\frac{W_j}{2}(Z_j-T^{col})\Big)\frac{\partial Y_j}{\partial \alpha_i}\\&+\frac{Y_jW_j\alpha_i}{2}(1+\alpha_j)^{1-n_j}\frac{\partial Z_j}{\partial \alpha_i}
\end{split}\end{equation*}
The derivatives $\frac{\partial W_i}{\partial \alpha_i}$, $\frac{\partial W_j}{\partial \alpha_i}$, $\frac{\partial Y_j}{\partial \alpha_i}$ and $\frac{\partial Z_j}{\partial \alpha_i}$ are respectively given by
\begin{equation*}\begin{split}
  \frac{\partial W_i}{\partial \alpha_i}=&-\frac{2}{\alpha_i}\Big((1+\alpha_i)\prod\limits_{j=0}^{N-1}(1+\alpha_j)^{-n_j}\Big)^{t_i-t_{min}+1}\cdot \\&\Big(\frac{1}{\alpha_i}+\frac{n_i-1}{1+\alpha_i}(t_i-t_{min}+1)\Big)
\end{split}\end{equation*}
\begin{equation*}\begin{split}
  \frac{\partial W_j}{\partial \alpha_i}=&-\frac{2n_i}{\alpha_j(1+\alpha_i)}(t_j-t_{min}+1)\Big((1+\alpha_j)\cdot \\&\prod\limits_{k=0}^{N-1}(1+\alpha_k)^{-n_k}\Big)^{t_j-t_{min}+1}
\end{split}\end{equation*}
\begin{equation*}
  \frac{\partial Y_j}{\partial \alpha_i}=-n_i(1+\alpha_i)^{-(n_i+1)}
\end{equation*}
\begin{equation*}
  \frac{\partial Z_j}{\partial \alpha_i}=(T_i^{succ}-T^{col})n_i
\end{equation*}
\subsubsection{Subgradient algorithm for optimal $\boldsymbol{\alpha}$}

Given the values of the Lagrange multipliers $\boldsymbol{\mu}^*$, the solution to Eqn.~(\ref{Lagrange}) specifies the optimal $\boldsymbol{\alpha}$. To complete the solution to the optimisation it therefore remains to calculate the optimal multipliers $\boldsymbol{\mu}^*$. These cannot be obtained in closed form since their values reflect the network topology. We proceed in a centralised manner by using a standard sub-gradient approach.
The dual problem for the primal problem defined in Eqn.~(\ref{opt}) is given by
\begin{equation*}
  \underset{\boldsymbol{\mu}\geq 0}
{\text{min}} \quad g(\boldsymbol{\mu})
\end{equation*}
where  the dual function $g(\boldsymbol{\mu})$ is given by
\begin{equation*}\begin{split}
  g(\boldsymbol{\mu})&=\underset{{\boldsymbol{\alpha}}}
{\text{max}}\ U'({ \boldsymbol{\alpha}})+\sum_{i=0}^{N-1}\mu_i(m_id_i-D_i(\boldsymbol{\alpha}))\\
&=U'\big({ \boldsymbol{\alpha}^*(\boldsymbol{\mu})}\big)+\sum_{i=0}^{N-1}\mu_i\big(m_id_i-D_i(\boldsymbol{\alpha}^*(\boldsymbol{\mu}))\big)
\end{split}\end{equation*}
For any $\boldsymbol{\alpha}$,
\begin{equation*}\begin{split}
   g(\boldsymbol{\mu})&\geq U'\big({ \boldsymbol{\alpha}}\big)+\sum_{i=0}^{N-1}\mu_i\big(m_id_i-D_i(\boldsymbol{\alpha})\big)
\end{split}\end{equation*}
and in particular, the dual function is larger than that for $\boldsymbol{\alpha}=\boldsymbol{\alpha}^*(\bar{\boldsymbol{\mu}})$, i.e.
\begin{equation*}\begin{split}
   g(\boldsymbol{\mu})&\geq U'\big({ \boldsymbol{\alpha}^*(\bar{\boldsymbol{\mu}})}\big)+\sum_{i=0}^{N-1}\mu_i\big(m_id_i-D_i(\boldsymbol{\alpha}^*(\bar{\boldsymbol{\mu}}))\big)\\
   &=g(\bar{\boldsymbol{\mu}})+\sum_{i=0}^{N-1}(\mu_i-\bar{\mu}_i)\big(m_id_i-D_i(\boldsymbol{\alpha}^*(\bar{\boldsymbol{\mu}}))\big)
\end{split}\end{equation*}
A sub-gradient of $g(\cdot)$ at any $\bar{\boldsymbol{\mu}}$ is thus given by the vector
\begin{equation*}
  \Big[m_id_i-D_i(\boldsymbol{\alpha}^*(\bar{\boldsymbol{\mu}}))\Big]_{i\in\{0,1,\cdots,N-1\}}
\end{equation*}
and the projected sub-gradient descent update is
\begin{equation*}
  \mu_i^{(t+1)}=\Big[\mu_i^{(t)}-\gamma\cdot \Big(m_id_i-D_i\big(\boldsymbol{\alpha}^*({\boldsymbol{\mu}}^{(t)})\big)\Big)\Big]^+
\end{equation*}
where $\gamma>0$ is a sufficiently small stepsize, and $[f(\cdot)]^+:=\max\{f(\cdot), 0\}$ ensures that the Lagrange multiplier never goes negative~\cite{Opt}.

The subgradient updates for $\boldsymbol{\mu}$ can be carried out centrally by the AP. For the $i$th AC, the AP requires the knowledge of the number of stations $n_i$, the PHY date rate $r$ and the packet size $L$. The algorithm to calculate optimal $\boldsymbol{\alpha}$ is detailed in Algorithm~\ref{alg}.
\begin{algorithm}
\caption{Calculate optimal $\boldsymbol{\alpha}$ }\label{alg}
\begin{algorithmic}[1]
\STATE Initialise
$\boldsymbol{\mu}^{(1)}=[\mu_0^{(1)},\mu_1^{(1)},\cdots,\mu_{N-1}^{(1)}]$,
$t=0$. \REPEAT \STATE  $t=t+1$; $\gamma^{(t)}=1/t^2$
\STATE  solve for $\boldsymbol{\alpha}^*(\boldsymbol{\mu}^{(t)})$ by combining equations $f_i(\boldsymbol{\alpha},\boldsymbol{\mu}^{(t)})=0 \qquad \forall i\in \{0,1,\cdots,N-1\}$
\STATE  $\forall i\in\{0,1,\cdots,N-1\}$,  calculate
\begin{align*}
 & \mu_i^{(t+1)}=\Big[\mu_i^{(t)}-\gamma^{(t)}\cdot \Big(m_id_i-D_i\big(\boldsymbol{\alpha}^*({\boldsymbol{\mu}}^{(t)})\big)\Big)\Big]^+
\end{align*}
\UNTIL
{\\(i) $\exists i\in \{0,1,\cdots,N-1\}, |m_id_i-D_i\big(\boldsymbol{\alpha}^*({\boldsymbol{\mu}}^{(t)})\big)|\leq{\epsilon}$, where $\epsilon>0$ and is sufficiently small; \\
(ii) $\forall i\in \{0,1,\cdots,N-1\}, m_id_i-D_i\big(\boldsymbol{\alpha}^*({\boldsymbol{\mu}}^{(t)})\big)\geq 0$ }.
\end{algorithmic}
\end{algorithm}

\section{Centralised closed-loop control approach}\label{Sec:ControlAlg}

In this section, we design a centralized adaptive control approach to implement the desirable proportional fairness in real networks.  Based upon the analysis in Section~\ref{Sec:Fairness}, the proportional fairness is achieved when the station attempt probability parameter $\boldsymbol{\alpha}$ reaches its optimum value $\boldsymbol{\alpha}^*$. The variable $\boldsymbol{\alpha}$ is only determined by the minimum contention window $W_i$ with $AIFS$ and $TXOP$ taking the recommended values and $CW_{max}=CW_{min}$. Our approach uses a multivariable closed-loop control system to tune $\boldsymbol{W}$  to drive the station attempt probability to its optimum. As the station attempt probability is hard to measure in real networks, we measure the conditional collision probability $\boldsymbol{p^{col}}(\boldsymbol{\tau})$ instead of $\boldsymbol{\alpha}(\boldsymbol{\tau})$ in the proposed control approach.

{
\subsection{Measuring $\boldsymbol{p^{col}}$}
The measuring of $\boldsymbol{p^{col}}$ is performed periodically at the AP every beacon interval. It requires messages passing from ordinary stations. {\color{black}A transmission from a station collides if a CTS is not received within a predetermined period after sending a RTS}.  Each station starts to count the number of transmitted RTS packets and that of missing CTS packets after receiving a beacon packet. {\color{black}The numbers counted are} piggybacked to the AP along with following RTS packets. After a beacon interval,  based on the piggybacked information, the AP {\color{black}knows} the number of transmitted RTS packets from stations of AC $i$, denoted as $N_i^{RTS}$, and the number of missing CTS packets of AC $i$, denoted as $N_i^{M-CTS}$, within a beacon interval. If we ignore the transmission duration of the beacon packet, and assume that packets collide with a constant collision probability within each interval, the average number of collided packets of AC $i$  is $N_i^{RTS}p_i^{col}$. These packets will be observed as missing CTSs, which yields $E(N_i^{M-CTS})=(N_i^{RTS})p_i^{col}$. Therefore,  we have
\begin{equation*}
    E\left(\frac{N_i^{M-CTS}}{N_i^{RTS}}\right)=\frac{(N_i^{RTS})p_i^{col}}{N_i^{RTS}}=p_i^{obs}
\end{equation*}

The observed conditional collision probability $p_i^{obs}$ can be estimated as $\frac{N_i^{M-CTS}}{N_i^{RTS}}$. To simplify notations, we hereafter refer to $\boldsymbol{p^{col}}$ with $\boldsymbol{p}$.
}

\subsection{Control algorithm}

\begin{figure}
  \begin{center}
  \includegraphics[width=0.5\textwidth]{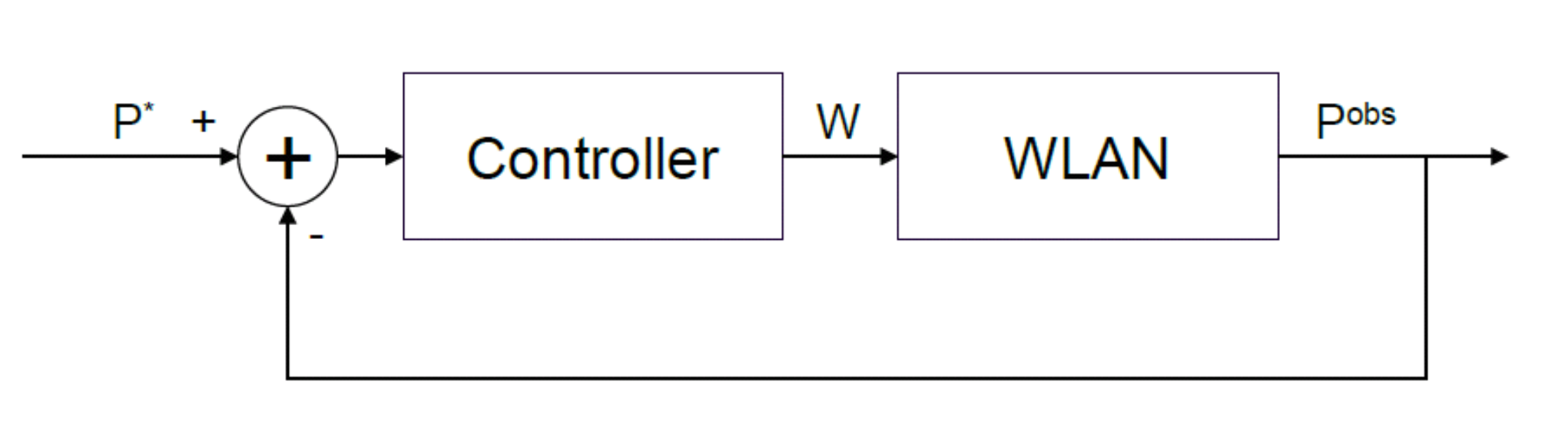}\\
   \caption{Closed-loop control system}\label{CtlSys}
  \end{center}
\end{figure}

The closed-loop control system consists of two modules as depicted in Fig.~\ref{CtlSys}.

The controller module is installed at the AP and  periodically carries out the adaptive algorithm every beacon interval, which is typically 100ms in 802.11 WLANs. {The adaptive control algorithm is described in Algorithm~\ref{Alg:control}:}

\begin{algorithm}
\caption{{Adaptive control algorithm}}\label{Alg:control}
\begin{algorithmic}[1]
{\STATE AP calculates the optimal $\boldsymbol{\alpha}^*$ and the resulting $\boldsymbol{p}^*$.
\STATE AP broadcasts the current $\boldsymbol{W}$ to all subscribed stations along with a beacon packet.
\STATE Each station starts to count the number of transmitted RTS packets and the number of missing CTS packets after receiving a beacon. The numbers are piggybacked to the AP along with following RTS packets.
\STATE At the end of a beacon interval (100ms), AP counts the total number of transmitted RTS packets of AC $i$, $N_i^{RTS}$, and the total number of missing CTS packets, $N_i^{M-CTS}$. The observed conditional collision probability $\boldsymbol{p^{obs}}$ in this interval is then calculated based on the counted numbers.
\STATE The reference $\boldsymbol{p}^*$ and the observed $\boldsymbol{p^{obs}}$ are input into the controller to calculate a new set of $\boldsymbol{W}$.
\STATE If the output of the controller $\boldsymbol{W}$ is not an integer, it is rounded to the closest integer value, and it has to be at least larger than 1. {\color{black}This requires hardware modification to allow CW to be chosen as an arbitrary integer.}
\STATE Go back to step 2.}
\end{algorithmic}
\end{algorithm}

  The plant is the WLAN itself.  The input of the plant is the contention window $\boldsymbol{W}=[ W_0,\cdots,W_{N-1}]$, and the output is the observed conditional collision probability $\boldsymbol{p^{obs}}=[p_0^{obs},\cdots,p_{N-1}^{obs}]$. The design objective is to obtain a stable system in closed-loop with desired performances and shape the output of the system to the given reference value. The reference value is the optimal conditional collision probability $\boldsymbol{p^*}$  given by Eqn.~(\ref{popt}) for $\boldsymbol{\tau}=\boldsymbol{\tau}^*$.

\subsection{Linearisation of the non-linear plant}\label{linearisation}

As the proposed adaptive control algorithm is executed every beacon interval, the period is long enough to assume that the measurement corresponds to stationary conditions.  This implies that $\boldsymbol{p^{obs}}$ depends only on the current $\boldsymbol{W}$, i.e. the system has no memory. Following this,
\begin{equation}\label{pvstau}
  p_i^{obs}=1-(1-\tau_i)^{n_i-1}\prod\limits_{\substack{j=0\\j\neq i}}^{N-1}(1-\tau_j)^{n_j}
\end{equation}
in which $\tau_i$ is a function of $W_i$, given by Eqn.~(\ref{CW}).

Eqn.~(\ref{pvstau}) and Eqn.~(\ref{CW}) give a non-linear relationship between $\boldsymbol{p^{obs}}$ and $\boldsymbol{W}$.
In order to simplify the controller design, we proceed by working with a linear approximation to this non-linear relationship around the stable point of operation.

The perturbations of input around the stable point of operation is
\begin{equation*}
  \boldsymbol{W}=\boldsymbol{W}^*+\delta\boldsymbol{ W}
\end{equation*}
in which $\boldsymbol{W}^*$ is the $\boldsymbol{W}$ which yields the optimal value $\boldsymbol{\tau}^*$ from Eqn.~(\ref{CW}).

The perturbations suffered by $\boldsymbol{p}^{obs}$ can be approximated by
\begin{equation*}
  \delta\boldsymbol{ p^{obs}}= \delta \boldsymbol{W}\cdot \boldsymbol{H}
\end{equation*}
in which
\begin{equation*}
  \boldsymbol{H}=\left(
     \begin{array}{cccc}
       \frac{\partial p^{obs}_0}{\partial W_0} & \frac{\partial p^{obs}_1}{\partial W_0} & \cdots & \frac{\partial p^{obs}_{N-1}}{\partial W_0} \\
       \frac{\partial p^{obs}_0}{\partial W_1} & \frac{\partial p^{obs}_1}{\partial W_1} & \cdots & \frac{\partial p^{obs}_{N-1}}{\partial W_{1}}  \\
       \vdots & \vdots & \ddots & \vdots \\
        \frac{\partial p^{obs}_{0}}{\partial W_{N-1}} & \frac{\partial p^{obs}_{1}}{\partial W_{N-1} }& \cdots & \frac{\partial p^{obs}_{N-1}}{\partial W_{N-1}} \\
     \end{array}
   \right)
\end{equation*}
The partial derivatives can be respectively calculated as
\begin{equation*}
   \frac{\partial p^{obs}_i}{\partial W_i}=\sum\limits_{k=0}^{N-1}\frac{\partial p^{obs}_i}{\partial \tau_k} \cdot \frac{\partial \tau_k}{\partial W_i}
\end{equation*}
and
\begin{equation*}
   \frac{\partial p^{obs}_i}{\partial W_j}=\sum\limits_{k=0}^{N-1}\frac{\partial p^{obs}_i}{\partial \tau_k} \cdot \frac{\partial \tau_k}{\partial W_j}
\end{equation*}
in which
\begin{equation*}
\frac{\partial p^{obs}_i}{\partial \tau_i}=\prod\limits_{k=0}^{N-1}(1-\tau_k)^{n_k}\frac{n_i-1}{(1-\tau_i)^2}
\end{equation*}
\begin{equation*}
\frac{\partial p^{obs}_i}{\partial \tau_j}=\prod\limits_{k=0}^{N-1}(1-\tau_k)^{n_k}\frac{n_j}{(1-\tau_i)(1-\tau_j)}
\end{equation*}
\begin{equation*}
  \frac{\partial \tau_i}{\partial W_i}=\frac{\tau_i^2}{-2\big(1-P_i^{blk}\big)\big({1}+(n_i-1)(t_i-t_{min}+1){\tau_i}\big)}
\end{equation*}
\begin{equation*}
  \frac{\partial \tau_i}{\partial W_j}=\frac{\tau_j(1-\tau_i)}{-2\big(1-P_j^{blk}\big)n_i(1-\tau_j)(t_j-t_{min}+1)}
\end{equation*}
At the stable point of operation, $\tau_i=\tau_i^*$, the non-linear plant is thus linearised as
\begin{equation}\label{output}
  \boldsymbol{p^{obs}}=\boldsymbol{W}\cdot \boldsymbol{H}(\boldsymbol{\tau^*})-\boldsymbol{W^*}\cdot \boldsymbol{H}(\boldsymbol{\tau^*})+\boldsymbol{p^{*}}
\end{equation}

 \subsection{State feedback control }

 With the lumbarisation, the WLAN can be represented as  a discrete MIMO LTI state-space model. According to the proposed adaptive algorithm, the conditional collision probability at instant $k+1$ is determined by the contention window input to the WLAN at instant $k$, the state and measurement equations are therefore given by
\begin{equation*}\left\{
                   \begin{array}{ll}
                     \boldsymbol{x}(k+1)=\boldsymbol{B}\boldsymbol{u}(k)\\
                     \boldsymbol{y}(k)=\boldsymbol{C}\boldsymbol{x}(k)
                   \end{array}
                 \right.
\end{equation*}
 in which the system state is the conditional collision probability,
\begin{equation*}
  \boldsymbol{x}(k)=[\boldsymbol{p^{obs}}(k)]^T
\end{equation*}
the system input is the minimum contention window,
\begin{equation*}
  \boldsymbol{u}(k)=[\boldsymbol{W}(k)]^T
\end{equation*}
and the system model matrices are
 \begin{equation*}
   \boldsymbol{B}=-\boldsymbol{H}^T
 \end{equation*}
  and
 \begin{equation*}
    \boldsymbol{C}=\boldsymbol{I}_{N\times N}
\end{equation*}
The system output is thus
\begin{equation*}
  \boldsymbol{y}(k)=\boldsymbol{C}\boldsymbol{x}(k)=[\boldsymbol{p^{obs}}(k)]^T
\end{equation*}

\begin{figure*}
  \begin{center}
  \includegraphics[width=0.8\textwidth]{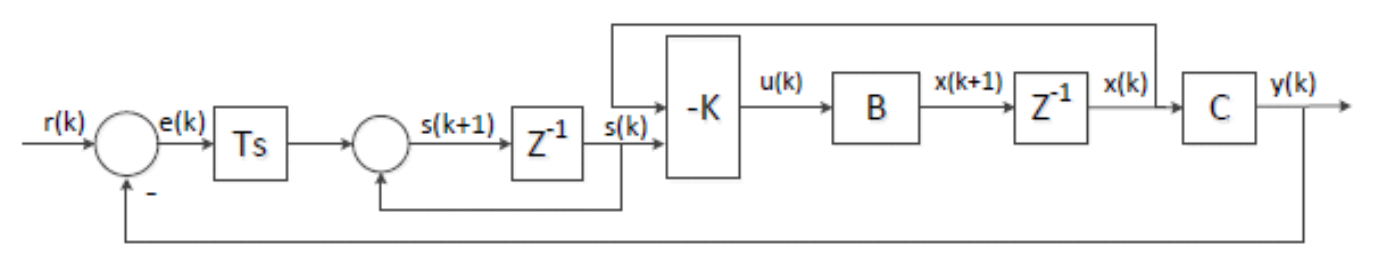}\\
   \caption{LQI Controller}\label{Fig：LQI}
  \end{center}
\end{figure*}

 The control task can be accomplished by using the LQI control method ~\cite{LQI} to design our controller. Fig.~\ref{Fig：LQI} shows the control block diagram for the system, in which $\boldsymbol{x}(k)\in \mathbb{R}^{N}$ is the system state, $\boldsymbol{y}(k)\in \mathbb{R}^{N}$ is the system output, $\boldsymbol{u}(k)\in \mathbb{R}^{N}$ is the controller output and $\boldsymbol{r}(k)\in \mathbb{R}^{N}$ is the controller input, which is the optimal collision probability $\boldsymbol{p^*}(k)$. $\boldsymbol{K} \in \mathbb{R}^{N\times 2N}$ is the control gain matrix, and $\boldsymbol{B} \in \mathbb{R}^{N\times N}$ and $\boldsymbol{C} \in \mathbb{R}^{N\times N}$ are the state-space system  matrices. $T_s$ is the sampling period of the system, i.e. the beacon interval 100ms.

The LQI controller computes an optimal state-feedback control law by minimising the quadratic cost function
\begin{equation*}
  J(\boldsymbol{u}(k))=\sum\limits_{k=0}^{\infty}\big(\boldsymbol{z}^T(k)\boldsymbol{Q}\boldsymbol{z}(k)+\boldsymbol{u}^T(k)\boldsymbol{R}\boldsymbol{u}(k)\big)
\end{equation*}
for any initial state $\boldsymbol{x}(0)$, in which $\boldsymbol{z}(k)=[\boldsymbol{x}(k);\boldsymbol{s}(k)]$ and $\boldsymbol{s}(k)=\boldsymbol{s}(k-1)+T_s \cdot (\boldsymbol{r}(k-1)-\boldsymbol{y}(k-1))$ is the output of a discrete integrator.

The matrices $\boldsymbol{Q}$ and $\boldsymbol{R}$ are the weighting matrices respectively indicating the state and control cost penalties. $\boldsymbol{Q}$ and $\boldsymbol{R}$ are required to be real symmetric and positive definite.

The state feedback control law is defined as
\begin{equation*}
  \boldsymbol{u}(k)=-\boldsymbol{K}\boldsymbol{z}(k)
\end{equation*}

The optimal state feedback gain matrix $\boldsymbol{K}$  is computed by solving the associated discrete algebraic Riccati equation
\begin{equation*}
  \boldsymbol{P}=\boldsymbol{\hat{A}^T}(\boldsymbol{P}-\boldsymbol{P}\boldsymbol{\hat{B}}(\boldsymbol{R}+\boldsymbol{\hat{B}^T}\boldsymbol{P}\boldsymbol{\hat{B})}^{-1}\boldsymbol{\hat{B}^T}\boldsymbol{P})\boldsymbol{\hat{A}}+\boldsymbol{Q}
\end{equation*}
in which
\begin{equation*}
  \boldsymbol{\hat{A}}=[\boldsymbol{0}_{N\times 2N};\ \ -\boldsymbol{C}*T_s \ \ \boldsymbol{I}_{N\times N}]
\end{equation*}
and
\begin{equation*}
  \boldsymbol{\hat{B}}=[\boldsymbol{B} ;\ \ \boldsymbol{0}_{N\times N}]
\end{equation*}
$\boldsymbol{K}$ is constructed from the solution of the above algebraic Riccati equation $\boldsymbol{P^*}$ and weighting matrices $\boldsymbol{Q}$ and $\boldsymbol{R}$, which is given by
\begin{equation*}
  \boldsymbol{K}=(\boldsymbol{R}+\boldsymbol{\hat{B}^T}\boldsymbol{P^*}\boldsymbol{\hat{B}})^{-1}\boldsymbol{\hat{B}^T}\boldsymbol{P^*}\boldsymbol{\hat{A}}
\end{equation*}

{
\subsection{Selection of $\boldsymbol{Q}$ and $\boldsymbol{R}$ }

The selection of weighting matrices $\boldsymbol{Q}$ and $\boldsymbol{R}$ affect the performance of the LQI controller. A simplified form using only 3 degrees of freedom is chosen in this work. The matrices are of the form
\begin{equation}\label{Eqn:Q}
 \boldsymbol{Q}= \begin{bmatrix}
q_1\boldsymbol{I}_N & \Large{0}\\[0.3em]
\Large{0}& q_2\boldsymbol{I}_N
\end{bmatrix}_{2N\times 2N}
\end{equation}
 and
 \begin{equation}\label{Eqn:R}
   \boldsymbol{R}=\rho \cdot \boldsymbol{I}_{N\times N}
 \end{equation}
 in which $q_1\in {R}^+$ is the weight for the state feedback cost; $q_2\in R^+$ is the weight for the integral feedback cost and $\rho\in {R}^+$ is the weight for the input cost.
}

\section{Performance evaluation}\label{Sec:Simulation}

\subsection{Throughput and delay performance}
The main objective of this work is to achieve proportional fair allocation of station throughputs while satisfying specific delay constraints of different ACs. To verify if the proposed fairness algorithm meets this objective, we first evaluate the throughput allocation and delay performance. The results are obtained using Matlab based on the throughput and delay analysis in Section~\ref{SubSec:throughput},~\ref{SubSec:delay} and the proportional fairness algorithm described in Section~\ref{SubSec:fairness}. The throughput and delay analysis is based on the 802.11e performance model presented in~\cite{Kosek} under the assumptions that stations have saturated traffic and $CW_{max}=CW_{min}$. The accuracy of this network model has been fully verified in~\cite{Kosek} for different network scenarios.

An example of an 802.11e WLAN with traffic of two ACs is considered, one of which is data traffic belonging to AC$\_$BE (best effort), and the other is video traffic belonging to AC$\_$VI (video). {Flows of both ACs are saturated.} The 802.11 OFDM PHY layer is assumed to be used. The PHY data rates for two ACs are the same, i.e. $r=54$Mbps.  The packet size is $L=8000$ bits. The 802.11 protocol parameters used in the evaluation are listed in Table~\ref{protocol_para}.
\begin{table}
\caption{802.11 protocol parameters used in the simulations}\label{protocol_para}
\centering
\begin{tabular}{|c|c|c|c|}
  \hline
   $\sigma$                       & 9  $\mu$s & $T_{PHYhdr}$          & 20  $\mu$s   \\ \hline
  $SIFS$                     & 16 $\mu$s   & $T_{RTS}$     & 46.67$\mu$s  \\    \hline
  $DIFS$                    & 34 $\mu$s  & $T_{CTS}$      & 38.67 $\mu$s   \\   \hline
  $EIFS$                                    & 88.67 $\mu$s  &$T_{ACK}$      & 38.67 $\mu$s\\  \hline
 \end{tabular}
\end{table}
The EDCA contention parameters recommended for 802.11 OFDM PHY layer are listed in Table~\ref{access_para}.
\begin{table}
\centering
\caption{EDCA channel contention parameters for 802.11 OFDM PHY}\label{access_para}
\begin{tabular}{|c|c|c|}
  \hline
  Access categories & AIFSN & Max TXOP \\ \hline
   Background (AC$\_$BK) & 7 & 0 \\
   Best Effort (AC$\_$BE) & 3 & 0 \\
   Video (AC$\_$VI) & 2 & 3.008ms \\
   Voice (AC$\_$VO) & 2 & 1.504ms \\ \hline
 \end{tabular}
\end{table}
Note that the IEEE 802.11e standard also provides recommended values for $CW_{min}$ and $CW_{max}$. As our control approach searches for the optimal contention window to achieve proportional fairness, the default $CW_{min}$ and $CW_{max}$ are not used in the proposed approach.

\begin{figure}
  \centering
  \subfigure[Throughput]{\label{Fig:fair1}\includegraphics[height=5cm, width=0.4\textwidth]{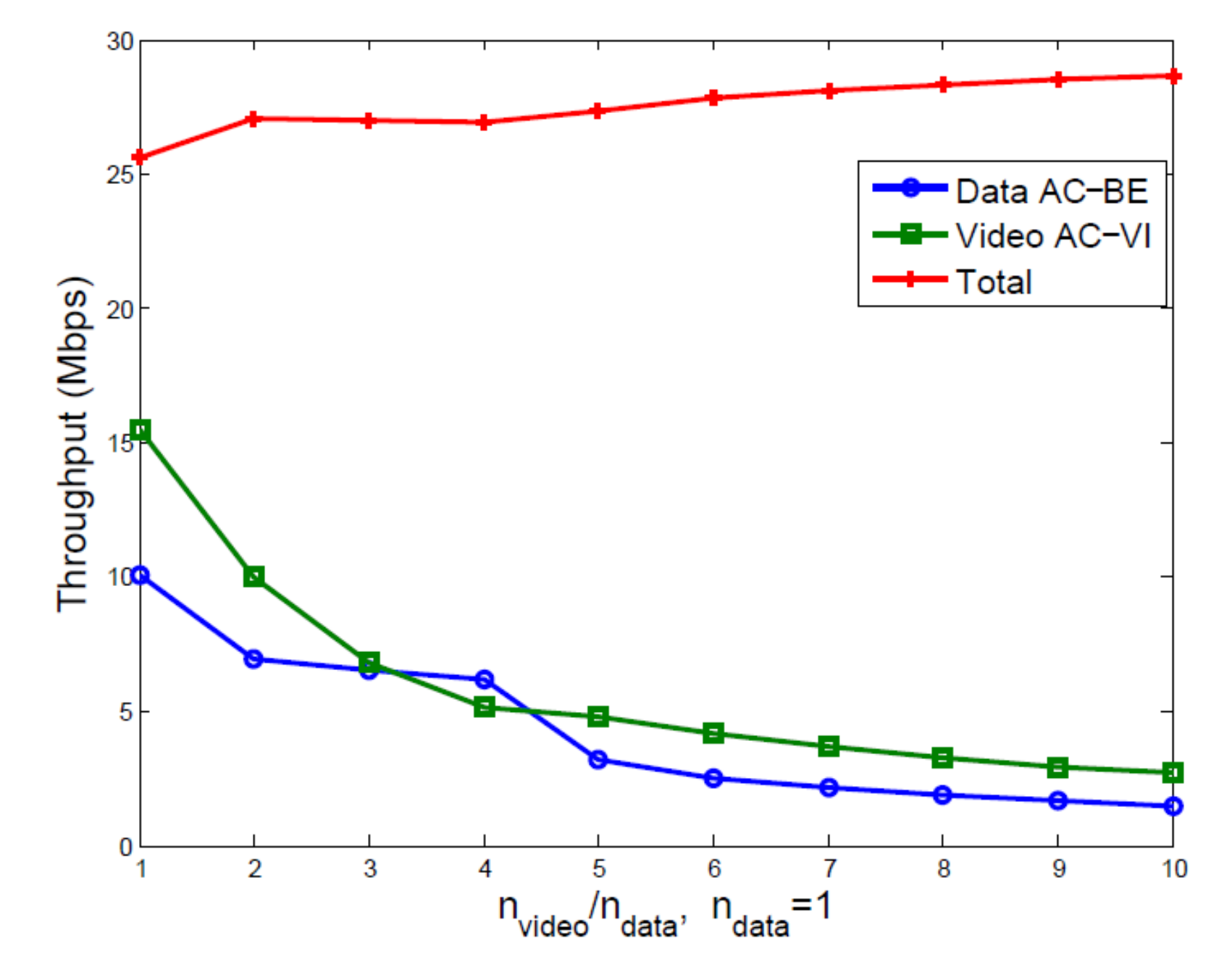}}
  \subfigure[Average delay]{\label{Fig:fair2}\includegraphics[height=5cm, width=0.4\textwidth]{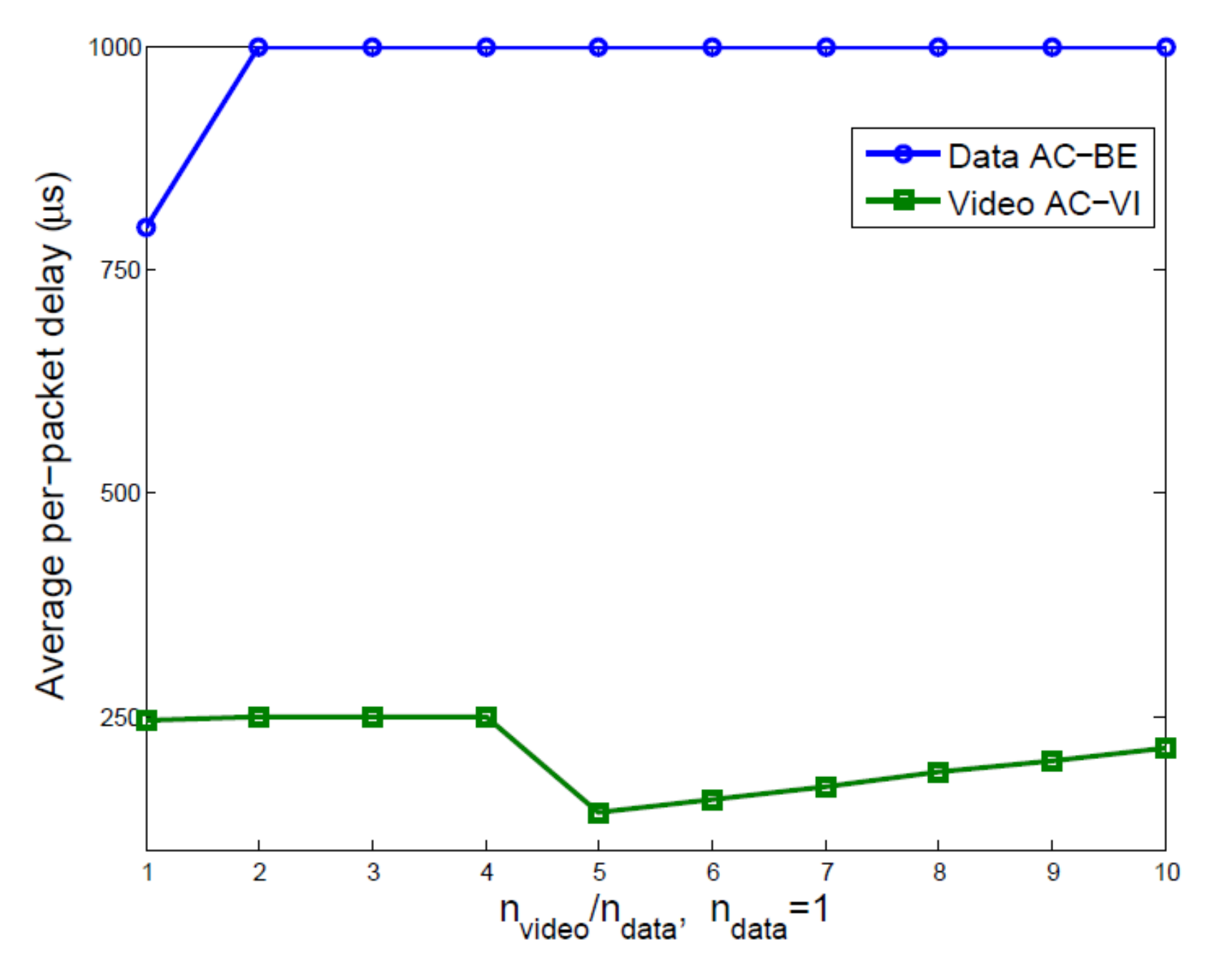}}
  \subfigure[Station attempt probability]{\label{Fig:fair3}\includegraphics[height=5cm, width=0.4\textwidth]{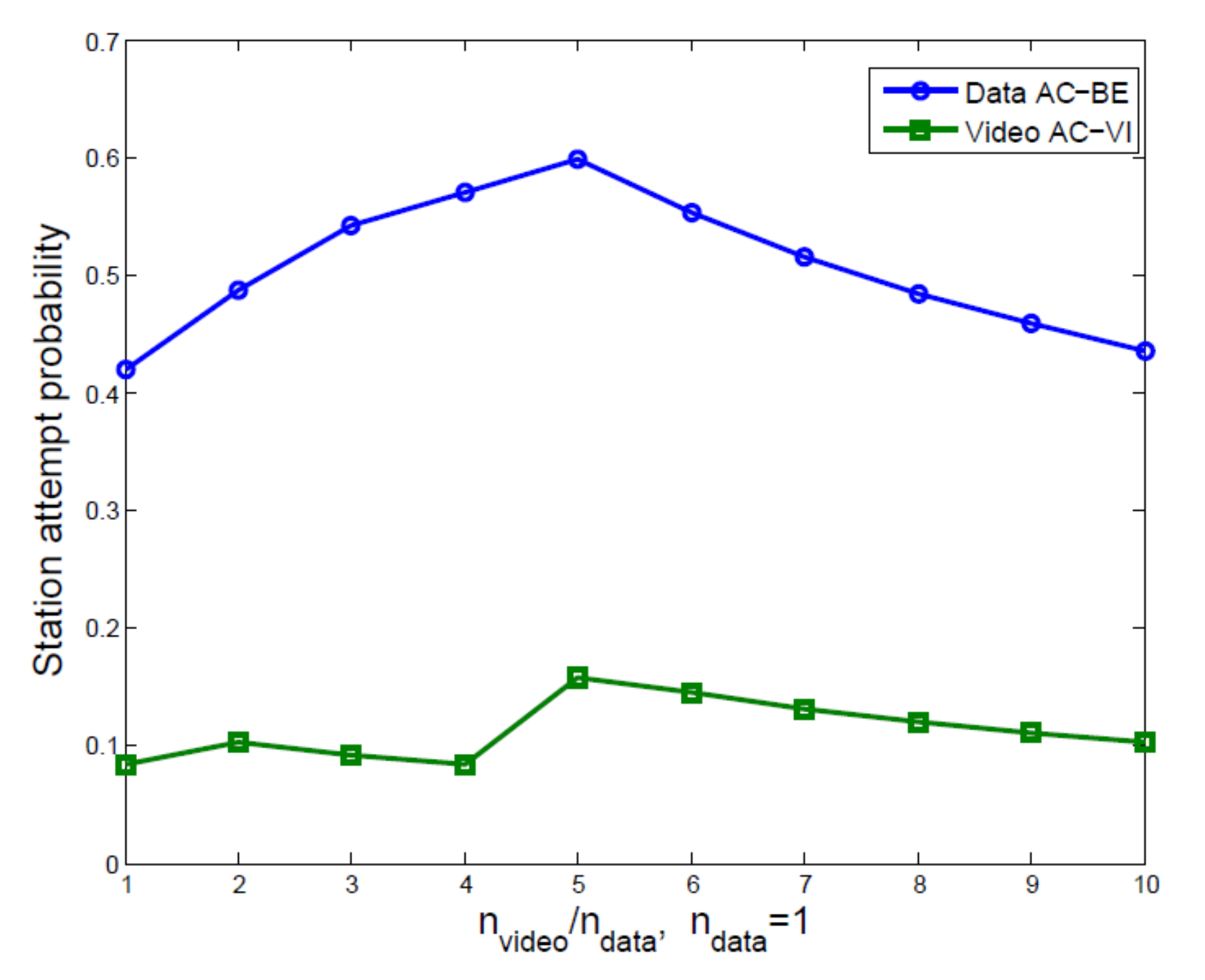}}
  \subfigure[{\color{black}Collision probability}]{\label{Fig:fair4}\includegraphics[height=6cm, width=0.4\textwidth]{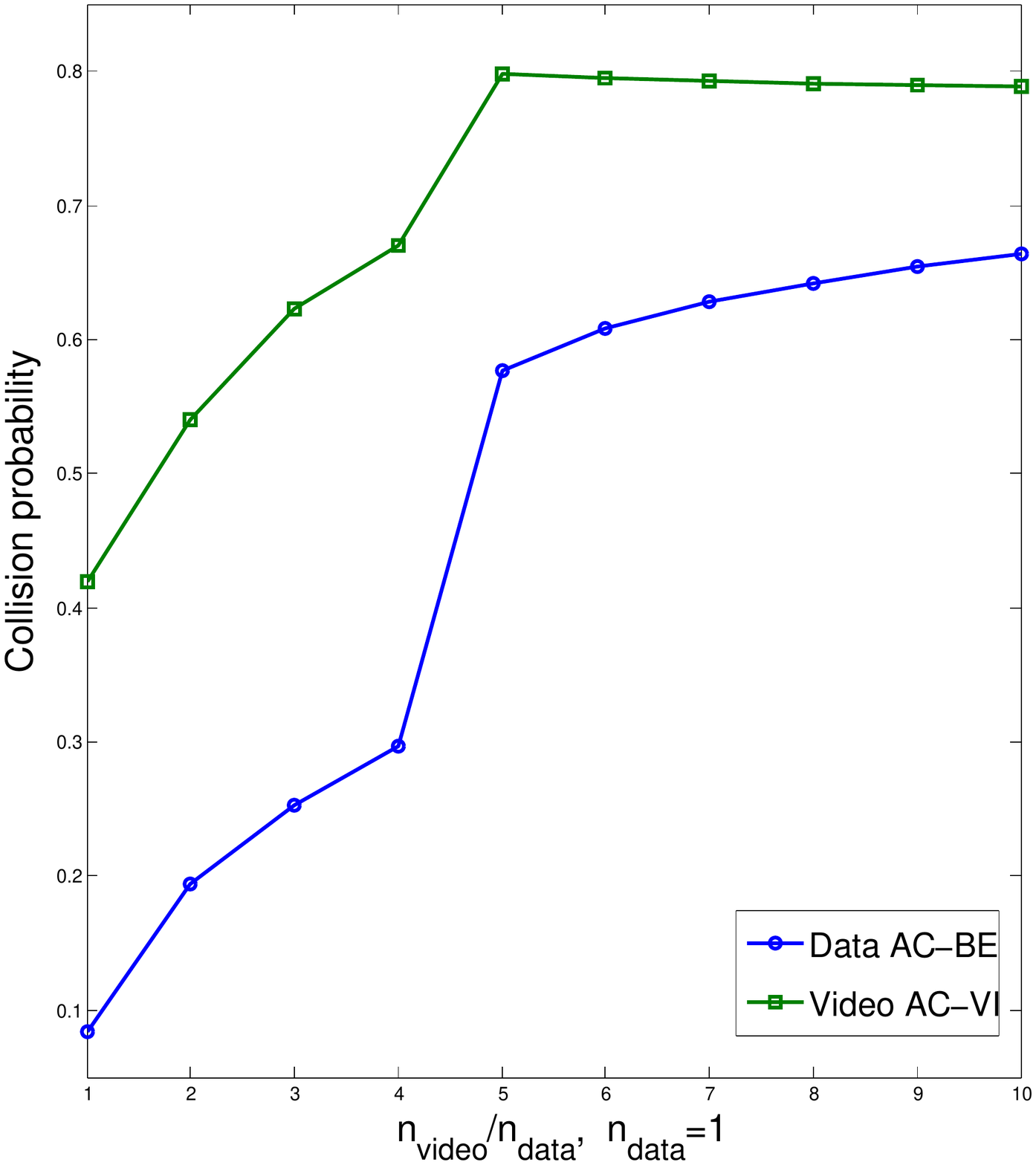}}
  \caption{Throughput and delay performances in an 802.11e WLAN with two ACs, the number of stations in AC$\_$BE $n_1=1$, $r=54$Mbps, $L=8000$bits, $d_1=1000\mu$s and $d_2=250\mu$s.}
\end{figure}

Fig.~\ref{Fig:fair1} and Fig.~\ref{Fig:fair2} shows the throughput and delay performance for two ACs versus the number of stations in AC$\_$VI while keeping the number of stations in AC$\_$BE fixed as 1. We let the TXOP burst reach the maximum limit as listed in Table~\ref{access_para}. The average delay deadline for a single video packet is $d_2=250\mu$s, which is the successful transmission duration of a video packet. The delay deadline for a single data packet is four times that of a video packet, i.e. $d_1=1000$$\mu$s. It can be seen that under these delay deadline constraints, the resource allocation can be divided into four phases. {Phase I: When there are one video station and one data station, the network load is quite light, and both delay deadlines are not reached yet. Phase II: When the number of video stations increases to 2, the increased collision possibility leads to longer delays, and so the delay deadlines of both ACs are reached. As the number of video stations increases, in order to achieve a fair throughput allocation, {\color{black}data stations attempt to access to the channel more} with an increased attempt probability, while video stations attempt less with a slightly decreasing attempt probability. The fairness algorithm makes the throughput of two ACs get closer to each other.   Phase III: When the number of video stations increases up to 5, it comes to the turning point when video traffic is so aggressive that the proportional fairness algorithm allocates higher throughput to video traffic by assigning increased attempt probabilities to both ACs, but the increase for video traffic is larger than that for data traffic. The delay of video traffic is then reduced to be less than the deadline limit, while the delay constraint of data traffic remains tight. Phase IV: As the number of video stations continues increasing, the throughput ratio between video and data traffic remains around 1.5 in this phase. Even with 10 video stations and only one data station,  the data station can still deliver a reasonable amount of throughput.} Fig.~\ref{Fig:fair3} shows the corresponding station attempt probabilities. It can be seen that contrary to the 802.11e EDCA standard, our algorithm assigns data traffic with a higher attempt probability although it has lower priority. However not only does the delay performance satisfy the QoS requirement, the throughput is also fairly allocated between two ACs. {\color{black}Fig.~\ref{Fig:fair4} shows the corresponding collision probabilities for both traffic types. As expected, the increased number of video stations results in an increase in collision probability in general. In Phase IV the collision probability for data traffic increases slowly, while that for video traffic slightly decreases.}

\subsection{Air-time}

The presence of collision losses and the coupling of station transmissions via carrier sense make the flow air-time in a WLAN not simply be the successful transmission duration but also include air-time expended in collisions. We define the \emph{flow total air-time} as the fraction of time used for transmitting a flow, including both successful transmissions and collisions. For a flow of AC $i$, the flow total air-time is
\begin{equation*}\begin{split}
  T_i^{air}&=\frac{P_i^{succ}T_i^{succ}+\tau_iP_i^{col}T^{col}}{{P^{idle}}\sigma+\sum\limits_{i=0}^{N-1}n_iP_i^{succ}T_i^{succ}+(1-P^{idle}-P^{succ})T^{col}}\\
  &=\frac{1}{X}\cdot \left(\alpha_i\Big(\frac{T_i^{succ}}{T^{col}}-1\Big)+\frac{\tau_i}{P^{idle}}\right)
\end{split}\end{equation*}

The work in~\cite{Xiaomin,Alex} finds that the proportional fair allocation assigns equal total air-time to each flow in a WLAN, and the air-times sum to unity. We investigate the air-time allocation by considering an 802.11e WLAN with four ACs, in which AC$\_$VI and AC$\_$VO have two stations, i.e. $n_2=n_3=2$, and AC$\_$BE and AC$\_$BK have one station, i.e. $n_1=n_4=1$. The PHY rates for four ACs are the same, i.e. $r=54$Mbps. The packet size is $L=8000$ bits. Table~\ref{air_time} compares the flow total air-time allocations under different delay deadline constraints. In Case I, the average per-packet delay deadline for the four ACs are respectively $d_1=900$$\mu$s, $d_2=300$$\mu$s, $d_3=250$$\mu$s and $d_4=1800$$\mu$s, while in Case II, the average delay deadlines are relaxed as $d_1=d_2=d_3=d_4=5000\mu$s.
\begin{table}\footnotesize
\centering
\caption{Comparison of flow total air-time allocations under different delay deadline constraints. }\label{air_time}
\begin{tabular}{c c c  c c c}
  \hline
   Flow &  1  &  {2 and 3}  &    {4 and  5}  &6  &  Sum \\ \hline
   AC   &BE& VI & VO  & BK & \\ \hline
   Case I & 0.1565 & 0.1530  & 0.1550  & 0.1562 & 0.9287\\ \hline
   Case II & 0.1667  & 0.1667  & 0.1667  & 0.1667 & 1\\ \hline
 \end{tabular}
\end{table}

It can be seen that different from observations in~\cite{Xiaomin,Alex}, the flow total air-time is not equalised in Case I, and the sum of air-times is less than 1. We also notice that in this case the delay constraints for voice traffic Flow 2 and Flow 3 are tight. Nevertheless when the delay constraints are relaxed in Case II, none of the delay deadline constraints is tight. Flows are now allocated with equal total air-times, and the air-times sum to 1. It is thus found that the air-time allocation in our algorithm is affected by the imposed delay constraints. With tight delay constraints, the proportional fair allocation assigns flows with the exact amount of air-time needed by each of them. Air-time resource in the network is not completely occupied in such a case. However with loose delay constraints, flows can occupy all the available air-time resource, and air-time is evenly distributed amongst flows in a network as discovered in previous work.  It is worth pointing out that since the flow air-time usage overlaps due to collisions, the flow total air-times summing to unity does not imply that the channel idle probability $P^{idle}=0$.

\begin{figure}
  \centering
  \subfigure[$q_2=2000$, $\rho=0.005$ ]{\label{Fig:QRtuning1}\includegraphics[height=5cm, width=0.4\textwidth]{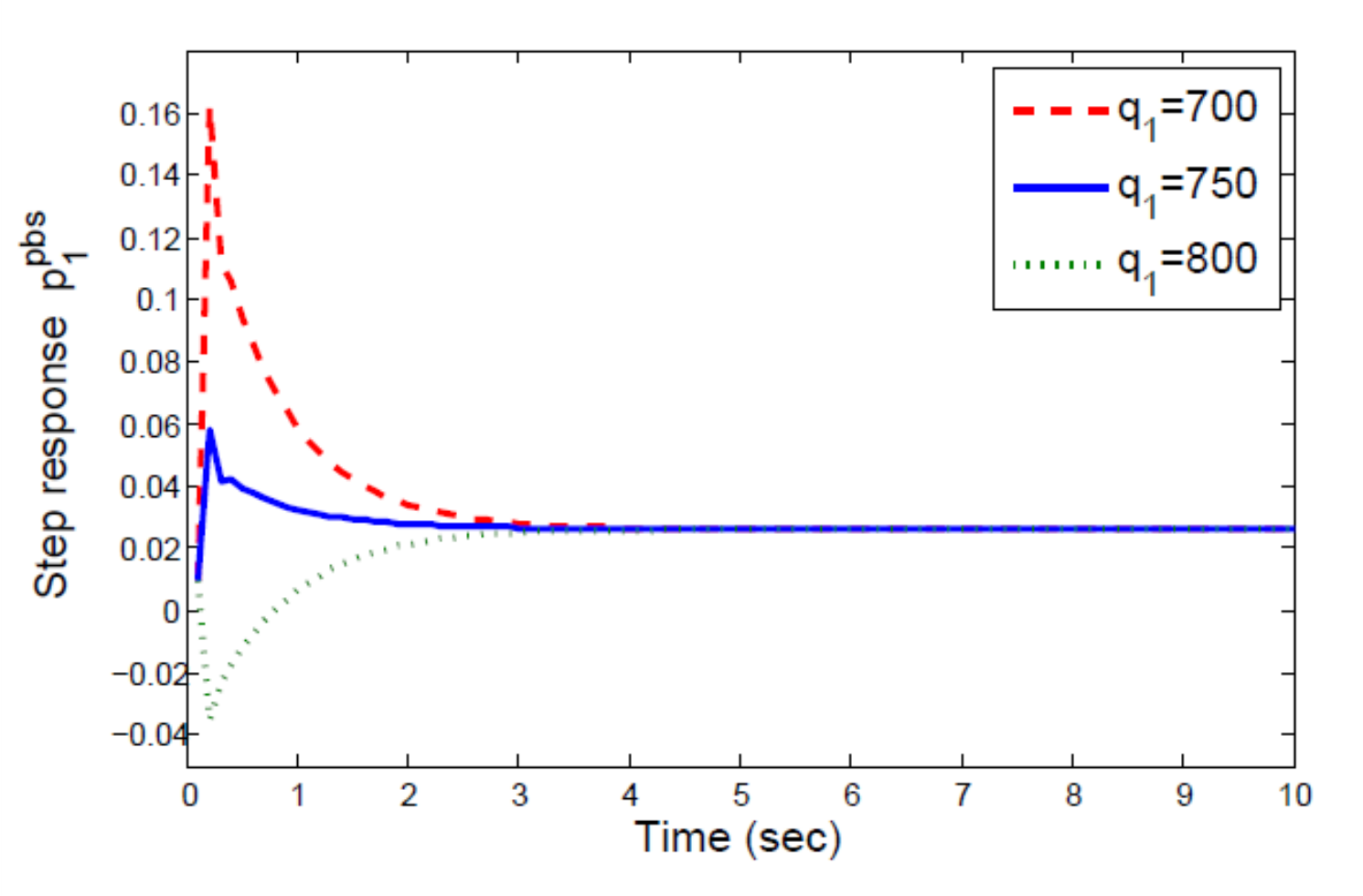}}
  \subfigure[$q_1=750$, $\rho=0.005$]{\label{Fig:QRtuning2}\includegraphics[height=5cm, width=0.4\textwidth]{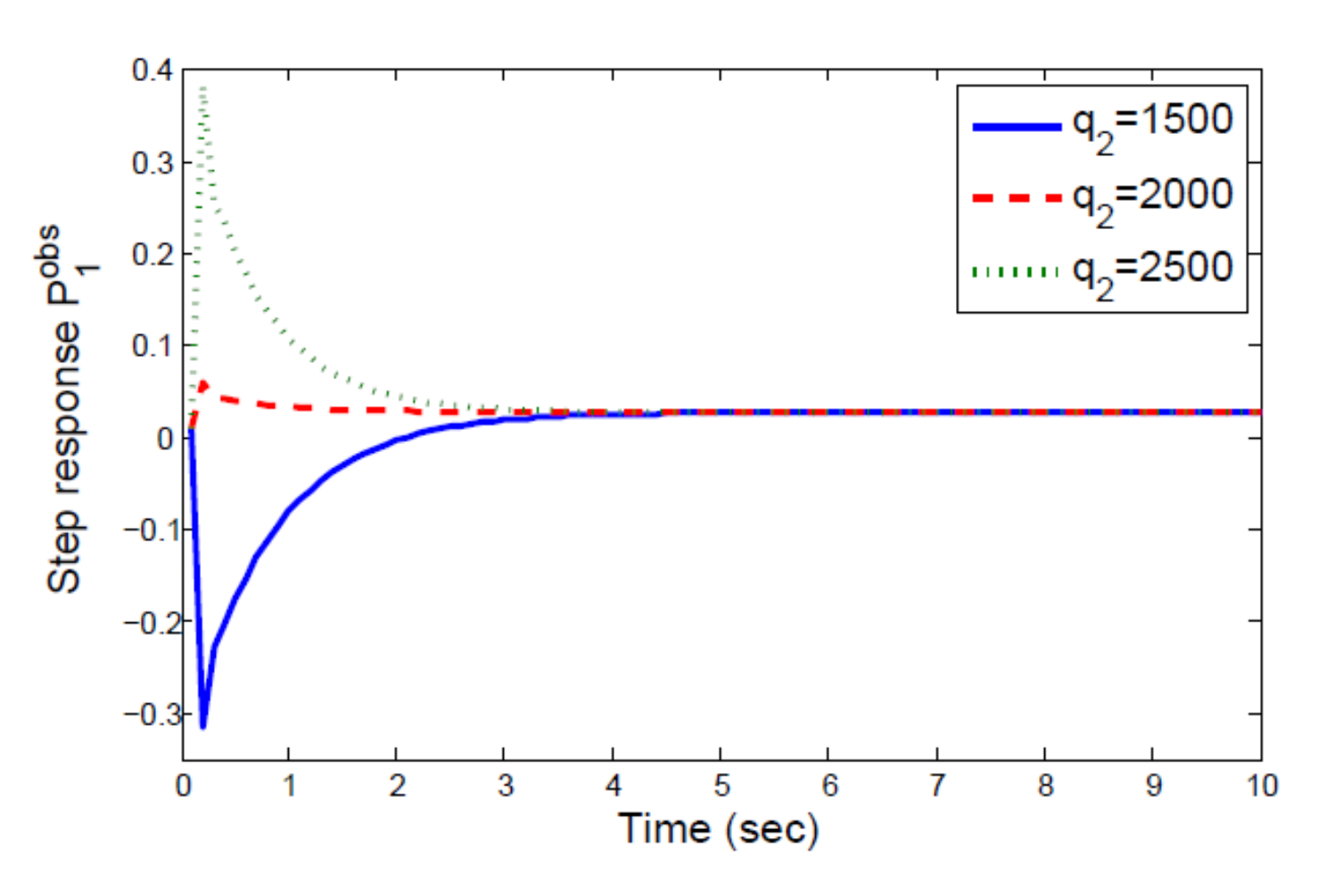}}
  \subfigure[$q_1=750$,$q_2=2000$]{\label{Fig:QRtuning3}\includegraphics[height=5cm, width=0.4\textwidth]{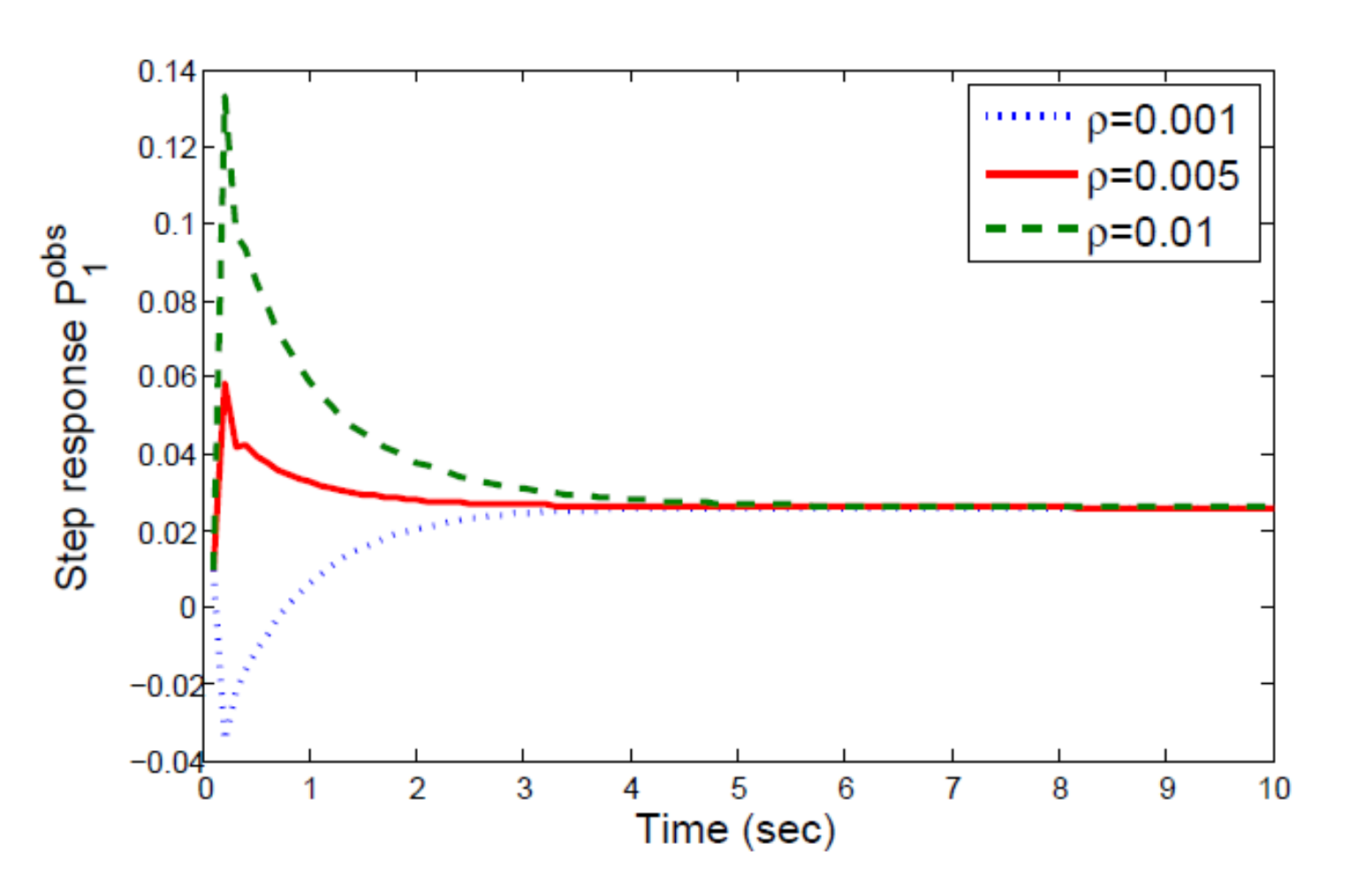}}
  \caption{System output step response, $n_1=1$, $n_2=2$, $n_3=1$, $r=54$Mbps, $L=8000$ bits, $d_1=900\mu$s, $d_2=300\mu$s $d_3=250\mu$s.}\label{Fig:QRtuning}
\end{figure}

{
\subsection{ $\boldsymbol{Q}$ and $\boldsymbol{R}$ tuning}

The tuning of $q_1$, $q_2$ and $\rho$ can be performed using {\color{black}the trial and error method}.  The influences of the three parameters are illustrated using an example with three ACs, AC$\_$BE, AC$\_$VI and AC$\_$VO. The number of stations in AC$\_$BE, AC$\_$VI and AC$\_$VO is respectively $n_1=1$,$n_2=2$ and $n_3=1$. Three ACs use the same PHY rate, i.e. $r_1=r_2=r_3=54$Mbps. The packet size is $l=8000$ bits. The average packet delay limit for AC$\_$BE, AC$\_$VI and AC$\_$VO is respectively $d_1=900\mu$s, $d_2=300\mu$s and $d_3=250\mu$s.  The TXOP burst reaches the maximum limit. Fig.~\ref{Fig:QRtuning} plots the system output response for AC$\_$BE with different sets of $q_1$, $q_2$ and $\rho$ values. The reason we do not put the output responses of AC$\_$VI and AC$\_$VO in Fig.~\ref{Fig:QRtuning} is that we notice that three outputs have the same convergence speed with a fixed set of $q_1$, $q_2$ and $\rho$ values. The effects of the three parameters are outlined as follows:
\begin{itemize}
  \item $q_1$ imposes the constraints to the state dynamics. It is directly related to the overshoot. A higher $q_1$ corresponds to a lower overshoot. As shown in Fig:~\ref{Fig:QRtuning1}, $q_1=700$ results in an overshoot, while $q_1=800$ corresponds to an undershoot.
  \item $q_2$ impacts on integral action dynamics and so on the system dynamics. As shown in Fig:~\ref{Fig:QRtuning2}, the higher it is, the smaller {\color{black}the} rising time will be, and the higher {\color{black}the} overshoot will be.
  \item $\rho$ affects the dynamics of the controller input, and so on the system dynamics. It is related to the overshoot.  A higher $\rho$ results in a higher overshoot, as shown in Fig:~\ref{Fig:QRtuning3}.
\end{itemize}
}

\subsection{Adaptivity to changes in the WLAN}

\begin{figure}
  \begin{center}
  \includegraphics[width=0.5\textwidth]{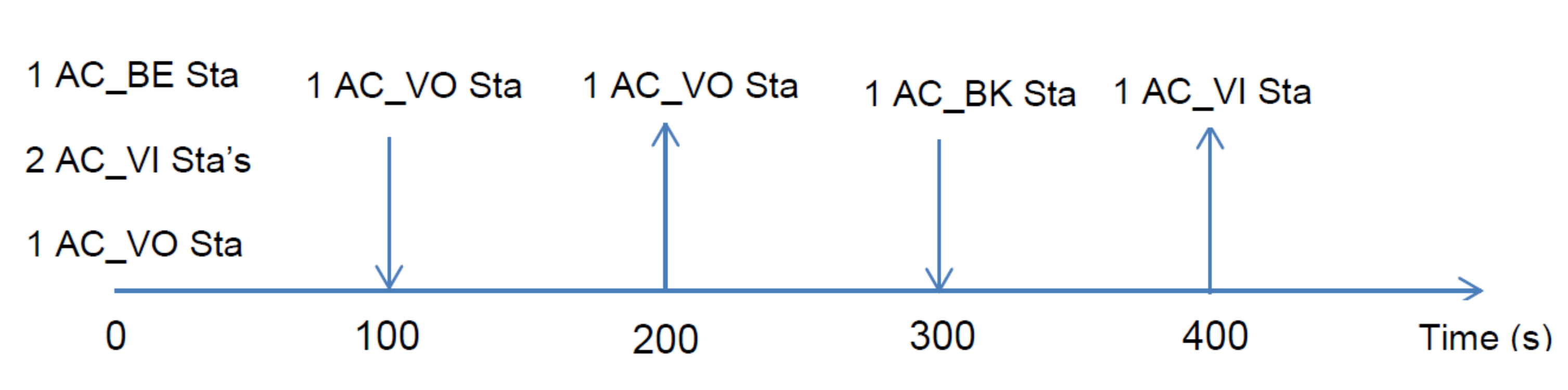}\\
   \caption{Injection and/or removal of stations in the WLAN. }\label{Fig:adaptivity}
  \end{center}
\end{figure}

We will next evaluate the adaptivity of the proposed method to changes in the network size. The scenario being considered is depicted in Fig.~\ref{Fig:adaptivity}. The algorithm starts at $t=0$s with four saturated stations, one in AC$\_$BE, two in AC$\_$VI and one in AC$\_$VO. One more station in AC$\_$VO joins the network at $t=100$s and leaves at $t=200$s. At $t=300$s one AC$\_$BK station joins the network, and after 100s one AC$\_$VI station leaves. The PHY data rates for the three ACs are the same, i.e. $r=54$Mbps. The packet size is $l=8000$ bits. The average packet delay limit for data, video, voice and background traffic are respectively $d_1=900\mu$s, $d_2=300$. $\mu$s, $d_3=250\mu$s and $d_4=1800\mu$s.
Fig.~\ref{Fig:CW} plots the variation of contention windows over time.
Fig.~\ref{Fig:thrpt} plots the corresponding station throughput for each AC. {\textbf{Q} and \textbf{R} take the form as displayed in Eqn.~(\ref{Eqn:Q}) and~(\ref{Eqn:R}). We choose $q_1=750$, $q_2=2000$ and $\rho=0.005$ to make a fast convergence speed.} It can be seen that when the network condition changes the contention window converges to the desirable value very quickly as long as proper {\textbf{Q} and {\textbf{R}} are chosen. Moreover, the steady-state errors can be neglected, which means the control system has high accuracy performance.
\begin{figure}
  \begin{center}
  \includegraphics[height=5.5cm, width=0.5\textwidth]{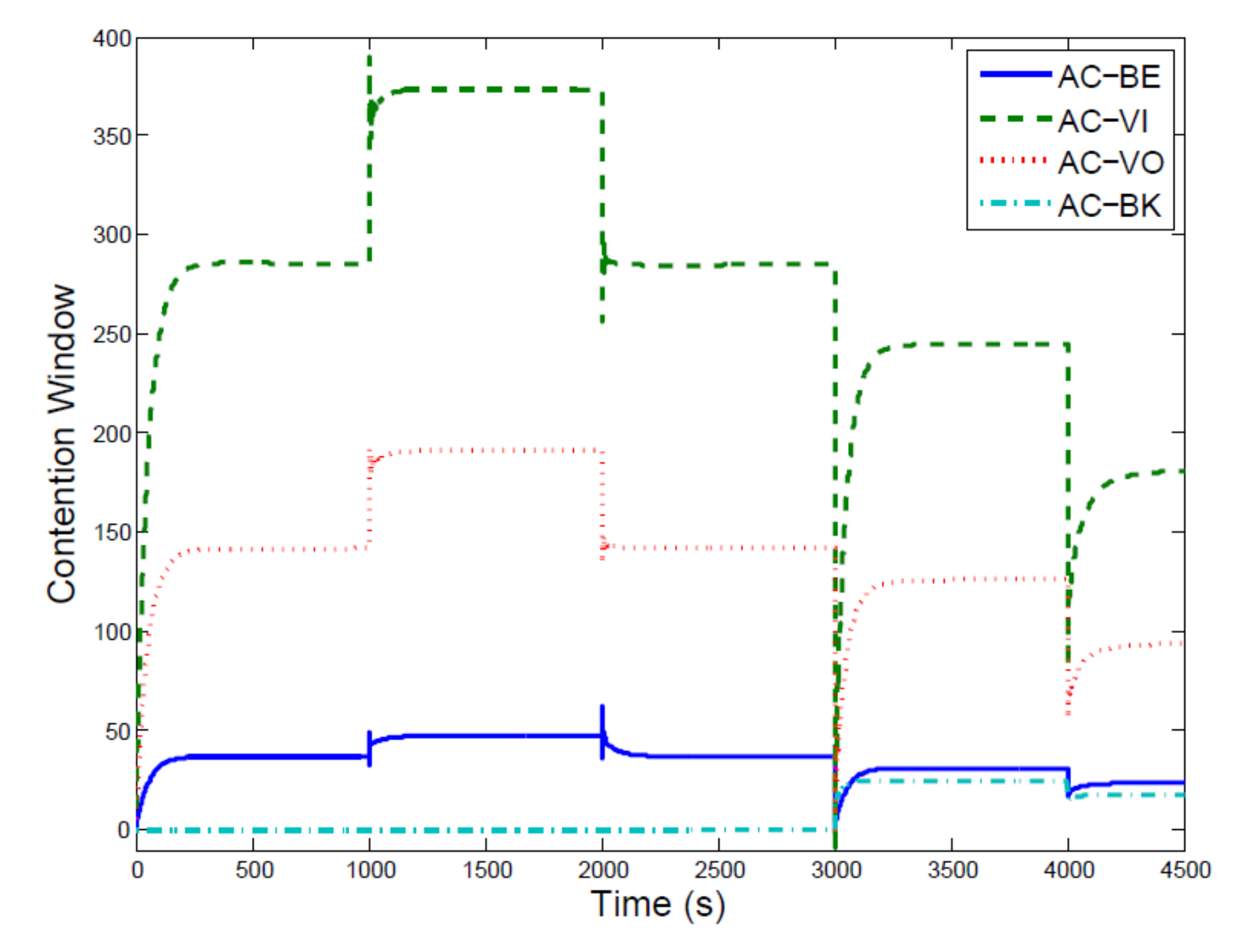}\\
   \caption{Contention window }\label{Fig:CW}
  \end{center}
\end{figure}
\begin{figure}
  \begin{center}
  \includegraphics[height=5.5cm,width=0.5\textwidth]{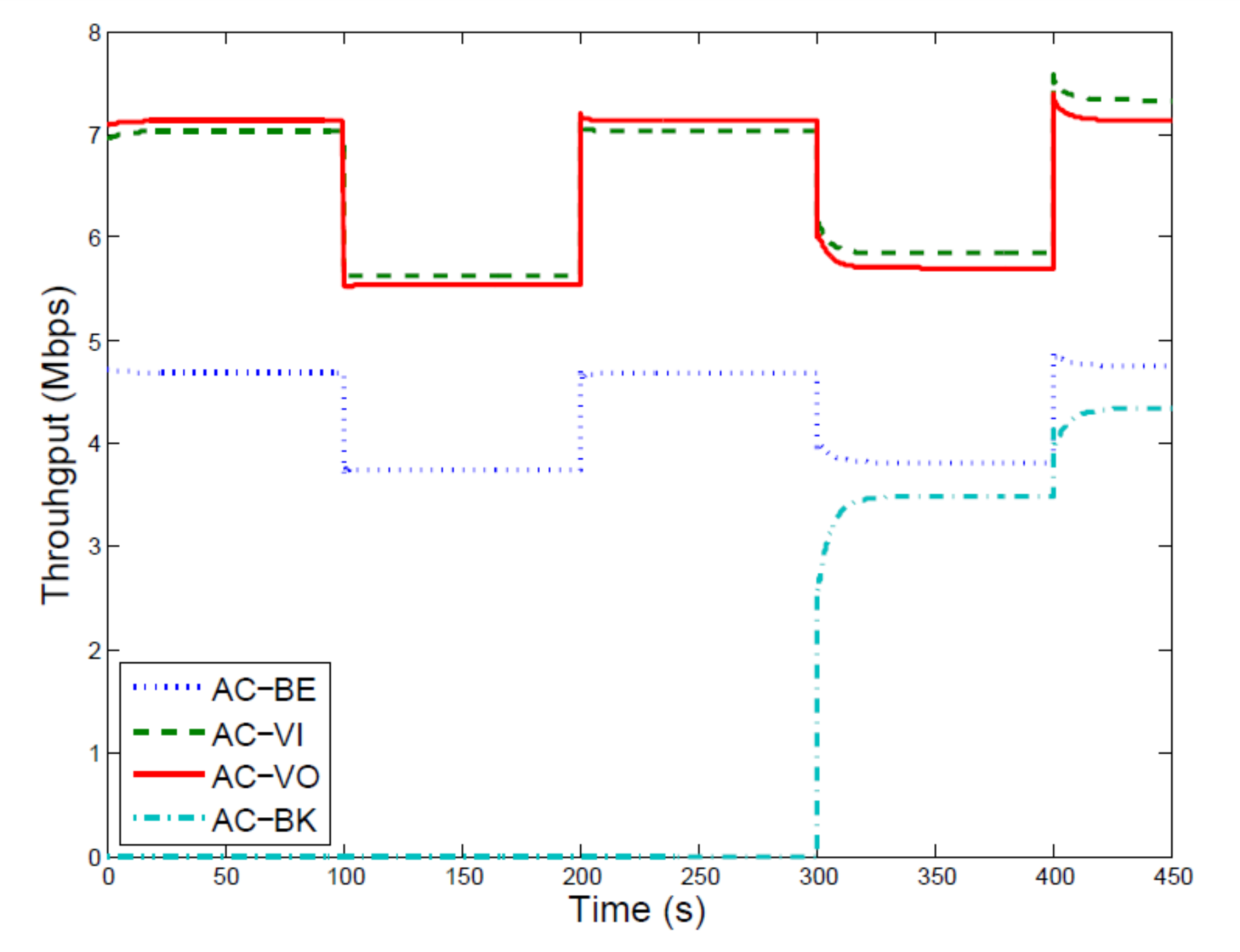}\\
   \caption{Station throughput }\label{Fig:thrpt}
  \end{center}
\end{figure}

\section{Conclusions}\label{Sec:Conclusion}
This paper considers using a closed-loop control approach to achieve proportional fair allocation of station throughputs in a multi-priority EDCA WLAN. The optimal station attempt probability that leads to proportional fairness is derived  given the average delay deadline constraints of different ACs present in an WLAN. To achieve the desirable proportional fairness, a centralised adaptive control approach is proposed. The WLAN is represented as a discrete MIMO LTI state-space model. The LQI control is used to tune the $CW_{min}$ value to the optimum. We have demonstrated using numerical results that the proposed control approach has high accuracy and fast convergence speed, and is adaptive to general network scenarios. To the best of our knowledge this might be the first detailed study of using a closed-loop control approach to achieve proportional fairness amongst ACs in EDCA WLANs. The optimisation of controller parameters is not considered in this paper. We leave that for future work.

\section{Acknowledgements}
This work was conducted as part of the project ISS-EWATUS (issewatus.eu) and has been financially funded by European Commission through the FP7 program (Grant agreement no:619228).

\appendix

\begin{proof}
\begin{equation*}
  U_1(\boldsymbol{\eta})=\sum\limits_{i=0}^{N-1}n_i\big(\eta_i-\log X(\boldsymbol{\eta})\big)
\end{equation*}
in which
\begin{equation*}\begin{split}
&\log X(\boldsymbol{\eta})\\&=\log \bigg(\frac{\sigma}{T^{col}}+\sum\limits_{i=0}^{N-1}n_i\Big(\frac{T_i^{succ}}{T^{col}}-1\Big)e^{\eta_i}+\prod\limits_{i=0}^{N-1}\big(1+e^{\eta_i}\big)^{n_i}-1\bigg)\\
&=\log \bigg(\frac{\sigma}{T^{col}}+\sum\limits_{i=0}^{N-1}n_i\frac{T_i^{succ}}{T^{col}}e^{\eta_i}+\prod\limits_{i=0}^{N-1}\big(1+e^{\eta_i}\big)^{n_i}-1-\sum\limits_{i=0}^{N-1}n_i e^{\eta_i}\bigg)\\
&=\log \bigg(\frac{\sigma}{T^{col}}+\sum\limits_{i=0}^{N-1}n_i\frac{T_i^{succ}}{T^{col}}e^{\eta_i}+\sum_{k=2}^{n}
\sum_{A\subseteq \N,|A|=k}\prod_{j\in A}{e^{\eta_j}}\bigg)
\end{split}\end{equation*}
and $\N=\{1,2,\cdots,n\}$ denotes the set of stations in the WLAN.

As the logarithm of a sum of exponentials is a convex function, $\log X$ is convex in the transformed variable $\boldsymbol{\eta}$, and $U_1(\boldsymbol{\eta})$ is thus concave in $\boldsymbol{\eta}$.
\end{proof}

\end{document}